%% file: main.tex
\documentclass[letterpaper,twocolumn,10pt]{article}
\usepackage{usenix-2020-09}
\input{sections/00preamble}
\begin{document}

\title{Lost at C: A User Study on the Security Implications\\  of Large Language Model Code Assistants}

\author{
{\rm \begin{NoHyper} Gustavo Sandoval\thanks{Equal Contribution}, Hammond Pearce\footnotemark[1], Teo Nys, Ramesh Karri, Siddharth Garg, Brendan Dolan-Gavitt\end{NoHyper}}\\
New York University
} %

\maketitle

\begin{abstract}
Large Language Models (LLMs) such as OpenAI Codex are increasingly being used as AI-based coding assistants. 
Understanding the impact of these tools on developers' code is paramount, especially as recent work showed that LLMs may suggest cybersecurity vulnerabilities.
We conduct a security-driven user study (N=58) to assess code written by student programmers when assisted by LLMs.
Given the potential severity of low-level bugs as well as their relative frequency in real-world projects, we tasked participants with implementing a singly-linked `shopping list' structure in C.
Our results indicate that the security impact in this setting (low-level C with pointer and array manipulations) is small: AI-assisted users produce critical security bugs at a rate no greater than 10\% more than the control, indicating the use of LLMs does not introduce new security risks. %

\end{abstract}

\input{sections/01Introduction}

\input{sections/02Background}

\input{sections/03Study}

\input{sections/04Results}

\input{sections/05Discussion}

\input{sections/09Conclusion}

\section*{Acknowledgments}
We gratefully acknowledge the participants from our user study, as well as OpenAI who provided API access to their \texttt{code-cushman-001} LLM for the duration.
This research was supported in part by NSF Grant \#2145482.

\Urlmuskip=0mu plus 1mu
\bibliographystyle{IEEEtran}
\bibliography{lit/user_study_auto}

\input{sections/10Appendix}

\end{document}

%% file: sections/00preamble.tex
\usepackage{graphicx}
\usepackage{textcomp}
\usepackage{xcolor}

\usepackage{booktabs}
\renewcommand*\descriptionlabel[1]{\hspace\leftmargin$#1$}

\def\BibTeX{{\rm B\kern-.05em{\sc i\kern-.025em b}\kern-.08em
    T\kern-.1667em\lower.7ex\hbox{E}\kern-.125emX}}
    
\usepackage[utf8]{inputenc}
\usepackage[most]{tcolorbox}

\usepackage{booktabs}

\usepackage{graphicx}
\usepackage{multirow}
\usepackage{setspace}
\usepackage{amsmath}
\usepackage{amssymb} %
\usepackage{amsfonts} %
\usepackage{calligra} %
\usepackage{mathtools}
\usepackage{amsthm} %
\usepackage{url}

\usepackage{dblfloatfix} %
\usepackage{array}
\usepackage[inline]{enumitem}

\usepackage{pgfplots}
\usepackage{pgfplotstable}
\usepgfplotslibrary{statistics}

\makeatletter
\let\MYcaption\@makecaption
\makeatother
\usepackage[font=footnotesize]{subcaption}
\makeatletter
\let\@makecaption\MYcaption
\makeatother

\usepackage{comment}
\usepackage{listings}

\usepackage{circledsteps}

\usetikzlibrary{patterns}

\usepackage{blindtext}

\usepackage{acronym}

\usepackage{xcolor, colortbl}

\usepackage{float}
\usepackage{textcomp}
\usepackage{algorithm}
\usepackage{algpseudocode}

\usepackage[all=normal, floats=tight,indent=tight]{savetrees}
\usepackage{pifont}
\usepackage{framed}
\usepackage{soul}
\usepackage{csquotes}
\usepackage{multirow}
\usepackage{array}
\newcolumntype{L}[1]{>{\raggedright\let\newline\\\arraybackslash\hspace{0pt}}m{#1}}
\newcolumntype{C}[1]{>{\centering\let\newline\\\arraybackslash\hspace{0pt}}m{#1}}
\newcolumntype{R}[1]{>{\raggedleft\let\newline\\\arraybackslash\hspace{0pt}}m{#1}}
\newcolumntype{H}{>{\collectcell\lstinline}l<{\endcollectcell}}
\MakeOuterQuote{"}

\newcommand{\newt}[1]{{{#1}}}

\definecolor{mygreen}{rgb}{0,0.6,0}
\definecolor{mygray}{rgb}{0.5,0.5,0.5}
\definecolor{mymauve}{rgb}{0.58,0,0.82}

\makeatletter
\let\old@lstKV@SwitchCases\lstKV@SwitchCases
\def\lstKV@SwitchCases#1#2#3{}
\makeatother
\usepackage{lstlinebgrd}
\makeatletter
\let\lstKV@SwitchCases\old@lstKV@SwitchCases

\lst@Key{numbers}{none}{%
    \def\lst@PlaceNumber{\lst@linebgrd}%
    \lstKV@SwitchCases{#1}%
    {none:\\%
     left:\def\lst@PlaceNumber{\llap{\normalfont
                \lst@numberstyle{\thelstnumber}\kern\lst@numbersep}\lst@linebgrd}\\%
     right:\def\lst@PlaceNumber{\rlap{\normalfont
                \kern\linewidth \kern\lst@numbersep
                \lst@numberstyle{\thelstnumber}}\lst@linebgrd}%
    }{\PackageError{Listings}{Numbers #1 unknown}\@ehc}}
\makeatother

\usepackage{lstlinebgrd} %
\usepackage{array}
\newcolumntype{L}{>{\centering\arraybackslash}m{3cm}}

%% file: sections/01Introduction.tex
\section{Introduction}

Large Language Models (LLMs) are deep neural networks trained on massive text corpora~\cite{brown_language_2020,radford_language_2019} to learn the underlying distribution of natural language or structured text.
When trained on code, LLMs can be used for code completion, bug fixing, and summarization~\cite{chen_evaluating_2021,ai21_discover_nodate, openai_examples_nodate}, useful features for developers.
Recent offerings are thus commercializing LLMs for code, including GitHub Copilot, which after its public release in June 2022 added 400,000 new users in just two months~\cite{torres_github_2022}.

However, recent work has shown that LLM completions may contain critical security vulnerabilities~\cite{siddiq_empirical_2022, pearce_asleep_2022}. 
This suggests that despite gain in developer productivity, LLM based code assistants 
should be used with caution (or not at all) due to security concerns.
This prior work has evaluated the security of LLM code assuming that its entirely generated by the LLM (we will call this the autopilot mode).
In practice, code completion LLMs \emph{assist} developers with suggestions that they can accept, edit or reject---a real-world security evaluation must account for the role of developers and how they interact with LLM based code assistants. 
While programmers prone to automation bias might naively accept buggy completions, other developers might produce overall less buggy code by only accepting safe suggestions and using time saved to fix other bugs.

\begin{figure}
    \centering
    \includegraphics{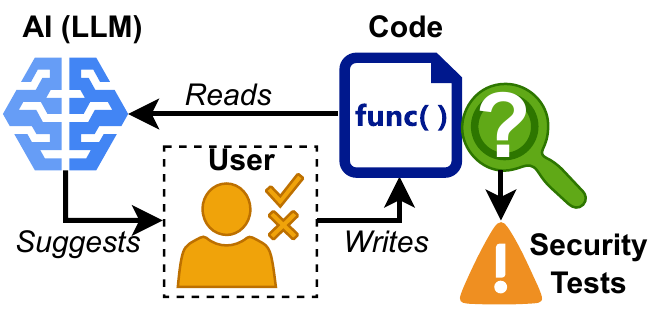}
    \caption{What is the security impact of LLM assistance?}
    \label{fig:general-question}
    \vspace{-4mm}
\end{figure}

This leads us to the key question motivating this work: 
\emph{Do developers with access to an LLM-based code completion assistants produce less secure code than the code produced by programmers without this access} (\autoref{fig:general-question})?
An affirmative answer to this question could be a significant showstopper for LLM based code assistants.
To answer this question, we perform the \emph{first} security-motivated randomized trial comparing programmers with and without access 
to a Codex-based code completion assistant powered by OpenAI's \texttt{code-cushman-001} LLM.

Our user study in \autoref{sec:study} had 58 computer science undergraduate and graduate students with programming backgrounds split randomly into `control' (no Codex LLM access) and `assisted' (with Codex LLM access) groups.
Given the relative frequency of memory-based errors in low-level languages such as C and C++ ($\approx$70\,\% of CWEs assigned by Microsoft each year~\cite{thomas_proactive_2019}), as well as their relative severity (classes of memory related bugs take many of MITRE's `Top 25 Common Weakness Enumeration (CWE) Most Dangerous Software Weaknesses' list, including positions \#1 and \#5~\cite{mitre_2022_2022}), we design a study where the participants were (t)asked to complete a set of 12 functions that perform basic operations on a linked list representing a ``shopping list'' in C.  
To understand the programming patterns in both groups, we created a cloud-based integrated development environment (IDE) that links to a Codex LLM in the back-end to behave like GitHub Copilot. 
The IDE logs user inputs and interactions with Codex LLM at a fine-grain. Using data from the user study, we investigate the impact of LLMs across three research questions:
\begin{enumerate}[label=\textbf{RQ\arabic*}:,leftmargin=1em,itemindent=2em,itemsep=0em]
\item Confirm  motivation for this research: Does an AI code assistant help novice users write better functional code? %
\item Given functional benefits, 
does the code that users write with AI assistance have acceptable incidence rate of security bugs vis-a-vis code written without assistance?
\item How do AI assisted users interact with ply vulnerable code suggestions---i.e., where do bugs originate in an LLM-assisted system?
\end{enumerate}

Our analysis, presented in \autoref{sec:results},  addresses these questions both quantitatively and qualitatively.
We examined completed code for functionality and security, using manual and automated methods. We sought to examine the code for bugs from the Common Weakness Enumeration (CWE)~\cite{mitre_cwe_2022} taxonomy.
We find that in our setting \newt{(a low-level linked list in C implemented by computer science students)}, the security impacts are minimal. We confirm existing findings on the productivity benefits of AI-assistance (RQ1), while finding that the AI-assisted group produced security-critical bugs at a rate no greater than 10\,\% higher than the control group (non-assisted) (RQ2).
When investigating the origin of bugs within the assisted users (RQ3), 63\,\% of the bugs originate in code written by humans and 36\,\% of the bugs were present in taken suggestions.
In the interests of open science we provide all data open source in~\cite{sandoval_lost_2022}.

%% file: sections/02Background.tex
\section{Background and related work}

\subsection{AI code assistants target productivity}

Academic and commercial AI code assistant tools are proliferating. Examples include OpenAI's Codex~\cite{chen_evaluating_2021,openai_examples_nodate}, AI21's Jurrasic J1~\cite{lieber_jurassic-1_2021,ai21_jurassic-1_2021}, Salesforce CodeGen~\cite{nijkamp_conversational_2022} and CodeBERT~\cite{feng_codebert_2020}.
These LLMs can write functionally correct code, with  %
 studies showing capabilities in solving introductory programming tasks~\cite{finnie-ansley_robots_2022,sarsa_automatic_2022} and algorithmic challenges~\cite{dakhel_github_2022}). %

Recent user studies examine the effects of code LLMs on developer productivity. 
Vaithilingam et al.~\cite{vaithilingam_expectation_2022} measured productivity from a group of developers (N=24) completing code tasks in Python. Each developer used GitHub Copilot to complete one task and the default non-AI based IntelliSense assistant to complete a different task and discussed which they preferred on three tasks of increasing difficulty. The exact task/assistant choice were randomized across participants. Overall, the participants preferred using Copilot as it helped them get started quicker. Analysis showed that the average task completion time when using Copilot was shorter although this result was not statistically significant---possibly because some participants did not complete the tasks in the allotted time, or due to the small sample size. 
They found Copilot generates code a lot quicker than typing or finding it from other sources. However, they also theorize that it is often buggy and so time saved writing code may then need to be spent in debugging Copilot generated code.

Imai~\cite{imai_is_2022} tasked a group of developers (N=21) to implement code for a `minesweeper' game. The study participants were randomly asked to use (1) GitHub Copilot as a code assistant, (2) a human pair programmer as the `driver' controlling the computer and writing the code, and (3) a human pair programmer as the `navigator' assisting `driver', and reading the code the `driver' is writing and examining it for issues. The study concluded that Copilot tended to result in more lines of code than with the human-based pair-programming in the same amount of time. However, the quality of code produced by Copilot was lower. Pair-programming with Copilot does not match the profile of human pair-programming. The study did not examine whether or not Copilot improves over a developer without a pair programmer.

A study by Ziegler et al.~\cite{ziegler_productivity_2022} from GitHub examines user perspectives on productivity during usage of GitHub Copilot.
Here, a large number of users running the GitHub Copilot technical preview were invited to complete a survey on their perspectives, and a subset of these responded (N=2,047).
Here, 84\% (1,724-out of-2,047 completions) self-scored a SPACE-type survey with a positive perspective aggregate. They felt Copilot had a more beneficial effect on their productivity than a negative one. 
Using internal metrics collected by the tool, GitHub authors determined that the number of suggestions accepted by the users was the greatest indicator of the positive perspective. The more suggestions a developer accepts, the more likely they feel the tool makes them productive. Github Copilot user's acceptance rate of suggestions is 6.6\% per hour.

A Google study by Tabachnyk et al.~\cite{tabachnyk_ml-enhanced_2022} 
used a large number of developers (N$>$10,000).
They found that since the deployment of a proprietary LLM, the fraction of all code added by the LLM has increased to 2.6\%, and developers have reduced their coding iteration duration by 6\,\% and reduced their number of context switches by 7\,\%, i.e. the LLM has had a measurable (and positive) impact on developer productivity.

\subsection{Prompts to suggestions: How LLMs code}
\label{sec:background:llms}
LLMs such as the GPT-type transformers which underpin Codex~\cite{chen_evaluating_2021} function by building probabilistic sequences of tokens based on the frequency of observed tokens in the training data~\cite{radford_language_2019}. In other words, they act as an `autocomplete' tool. 
Given some input sequence, they will find the most probable next token(s) in the output sequence. 
For instance, if an LLM is given ``\texttt{int main(int argc, char *}'' as the `prompt', 
it would likely return ``\texttt{argv}'' as the `suggested' next token, as this is a very common sequence in C programs.

In an LLM, `tokens' refer to common sets of individual characters.
These are used via `byte pair encoding'~\cite{gage_new_1994} to allow the LLMs to ingest more text into their fixed-size input windows. This allows the LLM to process more information.
Codex builds on the same tokenizer as GPT-3, extending it to include tokens for runs of whitespace. This makes it work better for code indentation~\cite{chen_evaluating_2021}. The average token for Codex is about four characters.

LLMs are not restricted to predicting just one token at a time, however.
They are autoregressive, feeding predictions back in on themselves and performing searches across chains of tokens (e.g. beam search is used in GPT-3~\cite{brown_language_2020}). In the code-writing LLMs, this allows for them to write large quantities of code `at once'. For example, given the right input prompt containing a well-defined function signature, an LLM may produce an entire function body.

\subsection{Security concerns of LLM-generated code}
\label{sec:bg-concerns}
Unfortunately, the somewhat na\"{i}ve mechanisms that underpin LLM suggestion generation discussed in the previous subsection have been shown to problematic outcomes from a security standpoint, for two primary reasons:
(1) LLMs may be trained over potentially insecure or buggy code (and will then reproduce those insecurities/bugs), and (2) Code which may be secure in isolation may be insecure depending on the sequence it is executed in relation to other pieces of code.

For an example of (1), consider the use of the `MD5' hash algorithm once widely used to protect secure information such as passwords. MD5 has been cryptographically broken, and so should no longer be used. However, code examples with MD5 remain on open source repositories. Therefore LLMs learn to (incorrectly) suggest MD5 for hashing passwords.
For (2), consider storing text in a buffer. This can occur safely using functions such as \texttt{snprintf}. However, if that buffer was just \texttt{free}-d, then the same line of code calling \texttt{snprintf} would result in a use-after-free vulnerability.

The issue of GitHub Copilot's code suggestions containing security vulnerabilities was first studied by Pearce et al.~\cite{pearce_asleep_2022}. They found that as measured by the GitHub CodeQL~\cite{github_inc_codeql_2021} static analysis tool, 
 40\,\% of the suggestions in relevant contexts contain security-related bugs (i.e. from MITRE's Common Weakness Enumeration (CWE) taxonomy~\cite{mitre_cwe_2022}).
Likewise, Siddiq et al.~\cite{siddiq_empirical_2022} showed that for the HumanEval dataset (which examine functional capabilities and not security) GitHub Copilot emits certain CWEs in around 2\,\% of cases as measured by Bandit analysis tool~\cite{bandit_developers_welcome_2022}.

That said, LLMs may also generate secure code as a replacement for insecure code~\cite{pearce_examining_2023} (i.e. may be used for bug patches). 
Although this was only shown to reliably work for small synthetic examples rather than for case studies taken from real-world vulnerabilities, where the results were inconclusive, this indicates that the issue of insecure code suggestions by LLMs remains unresolved.

Additional security concerns come from code beyond just their execution. The primary issue concerns code plagiarism~\cite{finnie-ansley_robots_2022}. 
Commercial organizations and academic institutions are worried that employees and students may misrepresent AI code LLM outputs as their own. Although outside the scope of this work, code licensing issues~\cite{ernst_ai-driven_2022,ciniselli_what_2022,topham_publication_2022} where the LLM may be trained over open-source licenses which impose requirements on their derivations. As the  legal status of code produced by LLMs is an open question, this could expose end-users to legal issues. %
Corporations may disallow employees from using public LLMs. On the other hand, in the university setting, it may be the case that using an AI  assistant could be academic dishonesty, especially if used in an examination setting (e.g. a solitary assignment). 
One example highlighting this was presented in \cite{biderman_neural_2022}, where students equipped with a code-writing GPT-J LLM passed introductory programming assignments without triggering suspicion from MOSS~\cite{aiken_system_2021}, a commonly-used anti-plagiarism software. In addition to the plagiarism-detection issue, they conclude that the LLMs will not be stumped with novel questions---GPT-J could solve problems outside the training set. 
When considering prose, rather than code, these issues have also been observed. Wahle et al.~\cite{wahle_are_2021} suggest using LLMs trained to detect plagiarism, and demonstrate that this technique can have greater success than typical tools such as TurnItIn. However, this has not yet been demonstrated for code.

\subsection{Evaluating code security}
\label{sec:background:subsec:codesec}

There are several techniques to determine the security of a piece of code. %
Identified bugs can be classified into the aforemntioned CWE taxonomy~\cite{mitre_cwe_2022}.

\textbf{Static Analysis} tools detect security-related bugs attempts statically at compile-time. Source code is parsed and analyzed for buggy design patterns. Common techniques include access-control analysis, information-flow analysis, and checks for application-programming-interface (API) conformance~\cite{pistoia_survey_2007}.
Common static analysis tools are listed by OWASP~\cite{owasp_source_nodate}, and include GitHub CodeQL~\cite{github_inc_codeql_2021}.

\textbf{Run-time analysis} can also occur by using tools such as debuggers and \textit{sanitizers} like `Address Sanitizer' (ASAN)~\cite{serebryany_addresssanitizer_2012} and `Undefined Behavior Sanitizer' (UBSAN) \cite{polacek_gcc_2014}.
These can identify bugs and instrument the underlying code at compile time to help identify the root cause, providing detailed information about the locations and causes of any errors. 
Unlike static analysis, sanitizers require a proof-of-concept `crashing' inputs which trigger bugs. These can be found by `fuzzers'.

\textbf{Fuzzers} run the program on concrete, randomly generated inputs in an attempt to uncover bugs and vulnerabilities. Bugs found with fuzzing are generally guaranteed to be true positives, and the proof-of-concept input that demonstrates the bug can be helpful to developers in fixes. Since the release of ``American fuzzy lop''~\cite{zalewski_american_2020} in 2013, fuzzing has received significant attention. Google's oss-fuzz~\cite{aizatsky_announcing_2016} provides continuous fuzzing for over 650 open source projects. 
Standards bodies such as NIST recommend fuzzing for secure development practices~\cite{black_guidelines_2021}. Major academic security conferences typically feature a dozen or more papers on fuzzing each year. While a full treatment is beyond the scope of this paper, we direct the reader to the survey by Manes et al.~\cite{manes_art_2021}. %

\textbf{Manual analysis:} Despite the pressing need for automated tooling, manual analysis for security bugs continues to be utilized at all stages of software design~\cite{tuma_automating_2020}. Manual code review is often essential to identify certain classes of bugs~\cite{di_biase_security_2016}. 
For example, in the study analysing Copilot's outputs~\cite{pearce_asleep_2022}, despite their use of GitHub CodeQL in identifying many CWE instances, other CWEs are checked manually.

In this study, we primarily use manual analysis jointly with static and run-time analysis as discussed in \autoref{sec:results:subsec:security}.

%% file: sections/03Study.tex
\section{Design of the security-focused user study}
\label{sec:study}

\subsection{Overview}
We seek to determine the impact of LLM assistance on the security qualities of the code written by programmers, comparing against baseline code written by programmers without assistance. 
Here, completions suggested by an LLM may be accepted by the users and inserted into their source code file. Suggestions may also be accepted and subsequently edited. %

Similar to the previous two user studies using LLMs \cite{imai_is_2022,vaithilingam_expectation_2022}, we examined undergraduate and postgraduate students from two software development courses at an R1 research university. We recruited participants by advertising on related social media. 
We tasked the participants with a programming assignment. 
Since real-world programming tends to be project-based (i.e. over a collection of related functions) rather than a collection of disparate tasks, we modeled the assignment in this manner. This is similar to Imai's~\cite{imai_is_2022} `minesweeeper' assignment.
We prompted participants to complete  a ``shopping list'' program implementation in C. 
This was chosen as a majority of bugs are memory-based issues in low-level languages such as C/C++~\cite{thomas_proactive_2019}.
Participants had to complete 12 functions related to this list (1 provided by us and 11 to complete). By providing a well-defined API (the list of functions), the program can be thought of as 12 separate programming tasks which may be analyzed separately. %
To minimize the risk of users running out of time, users were given two weeks to complete the assignment.

To understand the effects of the LLM suggestions, we randomly split the cohort into two groups.  The  `control' group were not given suggestions (the LLM was inactive).  The `assisted' group were given code suggestions by the LLM.   All groups were given identical video instructions to sign in to an online web portal where they complete the assignment in a controlled development environment, with an additional segment that either explains that they would get LLM suggestions and how to accept or reject them (`assisted' group), or that they would not get suggestions ( `control' group).  They were told that when they thought they were done, they should upload the program and complete an exit questionnaire for demographic information.
We analyzed the completed code for functional and security correctness.  This is discussed in \autoref{sec:results}.
Our institutional review board approved this study.

\subsection{Participant recruitment}
\label{sec:recruitment}

We recruit CS (or related discipline) students for our user study. Prior work has noted that CS students can be reasonable proxies for developers in the context of software engineering user studies.
Tahaei et al.~\cite{tahaei_recruiting_2022} found that ``recruiting CS students from our University’s mailing list resulted in the highest data quality in terms of programming skills (highest), costs (lowest), number of duplicates (low), and passing attention check questions (high) compared to the other tested crowdsourcing platforms.''
Further, Ko et al.~\cite{ko_practical_2015} note that university students can be appropriate participants when their knowledge and skills fit the one for the target audience. 
Finally,  Salman et al.~\cite{salman_are_2015} found that students and professionals do similarly on various code quality metrics. \newt{Intuitively, this makes sense: university level coding students will likely soon join industry as professional software developers.}

We selected participants with a range of experience from three sources:
(1) an undergraduate junior ``operating systems'' software class, (2) a senior- and MS-level ``Application Security'' software class, and (3) an informal student ``software chat group'' which operates over Discord app. During recruitment, we outlined the goals of this study to measure the impact of the LLM on code writing, and informed participants of a US\$50 compensation. \newt{For more details on the recruitment process and ethical considerations, see the Appendix.}

In total, 105 participants signed up for this study, %
and were randomly divided into the `assisted' group (to be prompted with code generated by the AI code assistant), and the `control' group (to not). \newt{The participants used their own computer and resources, and were informed that they could use the Internet to help in the general case, but they should not ask their peers or others for code writing assistance}. \newt{As noted in \cite{ko_practical_2015}, more structured studies in lab environments provide greater control at the expense of realism, and the choice between the two is subjective. Both types have been used in software-engineering literature.  Here, we opted for a more realistic setup at the expense of structure; as \cite{ko_practical_2015} notes, “if a tool is entirely new, it may be more valuable to observe a tool being used in more realistic conditions with fewer controls.” We expect the motivation levels between treatment and control to be similarly distributed because participants are randomly assigned to these groups.} Not all participants ended up engaging with the assignment: in total, 58 users completed code for analysis. We present demographics of these in \autoref{sec:demographics}.

\subsection{Programming assignment}
\textbf{Summary:}
The students were tasked with completing a programming assignment that consisted of a shopping list implemented using a singly linked list data structure%
\footnote{\url{https://en.wikipedia.org/wiki/Singly_linked_list}}. \newt{The complete assignment was provided all-at-once with all 11 functions simultaneously (task was to complete the C file to the specified API). Participants could implement functions in any order and update previous answers.}
The students were provided with documentation in the form of header files and a README, as well as an instructional video. 
Other supporting documents included a Makefile as well as 12 basic functional tests  such that the students could automatically test their code.

\textbf{Open source:} We open source the assignment in~\cite{sandoval_lost_2022}.

\textbf{Justification for the C Language:}
This study focuses on the \textit{security} implications of using LLM code assistants by programmers. 
Given the frequency of memory-related bugs in C\cite{thomas_proactive_2019} such as null pointer references, and array and buffer overflows, we chose the C programming language for the study; these bugs are both consequential (frequently leading to exploitable code and severe errors) and relatively easy for developers to inadvertently express via vulnerable design patterns. %
Further, unlike modern languages such as Go or Rust, the default compilation toolchains for C does not adequately check for these issues. %
Finally, all recruited participants should have experience programming in C due to syllabus requirements---i.e., this assignment should not be their first exposure to C.

\textbf{Instructions:}
Video and text instructions walked the students through signing in to the development environment, compiling, and  their code. %
We gave the students a folder that contained the files listed in \autoref{fig:study-docs}.
These contained documentation on what to do and how to do it (\texttt{README.md}, \texttt{Makefile}, \texttt{list.h}), the file they were to complete (\texttt{list.c}), and supplementary testing files to measure their progress and completion (\texttt{main.c}, \texttt{test.sh}, \texttt{unittests.exp}, \texttt{example\_load\_file.txt}).
These were designed similar to an industry-standard setup for programming - i.e., the users had to run \texttt{make test} to build and run tests and evaluate their code. %
Students were told they could use any resource on the Internet, such as Google and Stack Overflow. But they were not allowed to ask other students. We asked the students to complete as much of the functionality as they could during the two weeks of the study. They were not obligated to finish all functions to be compensated.  

\begin{figure}
    \centering
    \noindent\fbox{%
    \parbox{0.45\textwidth}{\texttt{%
    \scriptsize\textbf{* Core documents:}\\
README.md - contains study instructions\\
Makefile - script for compiling code/running tests\\
list.h - documentation and list API\\
list.c - file for participants to complete\\
\textbf{* Supplementary documents:}\\
main.c - instantiates a basic list application\\
runtests.c - the basic unit test suite \\
example\_load\_file.txt - for testing}}%
}
    \caption{Provided study documents / files}
    \label{fig:study-docs}
    \vspace{-4mm}
\end{figure}

\textbf{Introduction to the implementation:} 
The basic shopping list definition is provided in \autoref{fig:lst:node}. 
It is a singly-linked list with each node containing a \texttt{char*} string pointer, a price (float), and a quantity (int). 
No specific information is provided regarding other properties of these variables.
The users are then provided with the function APIs in the remainder of \texttt{list.h}, as \texttt{\#include}s and implementation hints at the beginning of \texttt{list.c} (\autoref{fig:lst:hints}).

\begin{figure}
\begin{subfigure}[b]{\linewidth}
\centering
\begin{lstlisting}[language=c]
// Node of the singly linked list
typedef struct _node {
    char* item_name;
    float price;
    int quantity;
    struct _node *next;
} node;
\end{lstlisting}
\vspace{-3mm}
\caption{Node definition (in \texttt{list.h})}
\label{fig:lst:node}
\end{subfigure}

\begin{subfigure}[b]{\linewidth}
\centering
\begin{lstlisting}[language=c]
#include <stdio.h> 
#include <stdlib.h>
#include <getopt.h>
#include <string.h>
#include "list.h"

#define MAX_ITEM_PRINT_LEN 100

// Note: All list_ functions should return a status code
// EXIT_FAILURE or EXIT_SUCCESS to indicate success.
\end{lstlisting}
\vspace{-3mm}
\caption{\texttt{\#include}s and implementation hints (in \texttt{list.c})}
\label{fig:lst:hints}
\end{subfigure}

    \vspace{-2mm}
\caption{Preliminary codes}
    \vspace{-4mm}
\end{figure}

\textbf{Basic functions:} The `basic' API functions are presented in \autoref{fig:lst:basic-list}. 
Here, `basic' refers not to the difficulty of the underlying code, but of the fundamental code required in any linked list implementation -- functions to add an item (at a position), update an item, remove an item, and swap two items. 
We chose to complicate matters by (1) making all API functions use a position index rather than importing and exporting node pointers; (2) making the linked-list \textbf{one}-indexed, rather than the more standard zero-indexed; and (3) making all I/O to the API via function arguments and argument pointers with the API functions instead of returning success/failure status.
These design choices increase the chance of unintended bugs, as they increase the complexity of traversing the linked list. %

\begin{figure}
\centering
\begin{lstlisting}[language=c]
// create a new list
int list_init(node **head);

// add a new item (name, price, quantity) to the list at position pos
int list_add_item_at_pos(node **head, char *item_name, float price, int quantity, unsigned int pos);

// update the item at position pos
int list_update_item_at_pos(node **head, char *item_name, float price, int quantity, unsigned int pos);

// remove the item at position *pos*
int list_remove_item_at_pos(node **head, int pos);

// swap the item at position pos1 with the item at position pos2
int list_swap_item_positions(node **head, int pos1, int pos2);
\end{lstlisting}
\vspace{-3mm}
\caption{Basic linked list functionality (comments simplified)}
\label{fig:lst:basic-list}
\end{figure}

\textbf{`Tricky' functions - string manipulation, advanced traversal, saving and loading:} The list of  complex, non-standard functions are presented in \autoref{fig:lst:advanced-list}. These are separated into print functions (\autoref{fig:lst:print-list}), the advanced traversal functions (\autoref{fig:lst:advanced-list}), and saving and loading functions (\autoref{fig:lst:saving-list}).
These functions also require index positions and argument pointers. Using pointers for return values increases the number of pointer manipulations needed for functionality, increasing the chance that code may be written with unintended security-relevant bugs. 
For example, \texttt{list\_item\_to\_string}, which uses a documented ``externally allocated string'', needs to be written to avoid potential buffer overflows (what is the maximum length this string is allowed to be? It is provided in the earlier \texttt{\#define} from \autoref{fig:lst:hints}). This is easy to miss. Other complex implementations arise from \texttt{list\_deduplicate}, which requires traversal and item removal code, and \texttt{list\_save} and \texttt{list\_load} which need to deal with files and the list. %

\begin{figure}
\begin{subfigure}[b]{\linewidth}
\centering
\begin{lstlisting}[language=c]
// print a single list item to an externally allocated string
// This should be in the format of:
// "quantity * item_name @ $x.xx price ea"
int list_item_to_string(node *head, char *str);

// print the list to stdout
// This should be in the format of:
// "pos: quantity * item_name @ $x.xx price ea" places.
int list_print(node *head);
\end{lstlisting}
\vspace{-3mm}
\caption{String manipulation functions (comments simplified)}
\label{fig:lst:print-list}
\end{subfigure}

\begin{subfigure}[b]{\linewidth}
\centering
\begin{lstlisting}[language=c]
// find the item position with the highest single price
int list_find_highest_price_item_position(node *head, int *pos);

// calculate the total cost of the list (sum of all prices * quantities)
int list_cost_sum(node *head, float *total);

// de-duplicate the list by combining items with the same name 
//    by adding their quantities
// The order of the returned list is undefined and may be in any order
int list_deduplicate(node **head);
\end{lstlisting}
\vspace{-3mm}
\caption{Advanced traversal functions (comments simplified)}
\label{fig:lst:advanced-list}
\end{subfigure}

\begin{subfigure}[b]{\linewidth}
\centering
\begin{lstlisting}[language=c]
// save the list to file filename
// the file should be in the following format:
// item_name,price,quantity\n 
int list_save(node *head, char *filename);

// load the list from file filename
// the file should be in the following format:
// item_name,price,quantity\n 
int list_load(node **head, char *filename);
\end{lstlisting}
\vspace{-3mm}
\caption{Saving and Loading the list (comments simplified)}
\label{fig:lst:saving-list}
\end{subfigure}
\vspace{-2mm}
\caption{Advanced implementation requirements}
\vspace{-4mm}
\end{figure}

\subsection{The `Codex Assistant' for code suggestions}
\label{sec:ide}
\begin{figure}
  \centering
  \fbox{\includegraphics[width=0.98\linewidth]{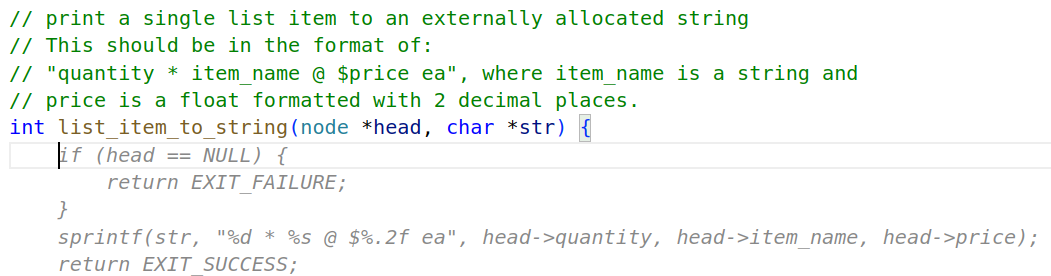}}
\caption{Example suggestion by Codex Assistant. Suggested code is in grey italic. Prompt is all text before the cursor.}
\label{fig:assistant-ex}
\end{figure}

In this section we will introduce the AI-based system which generates code for the `assisted' study group. 
This assistant was modelled after the commercial GitHub Copilot.
It is built as an extension for Visual Studio Code which parses the file under development, sends data to the OpenAI Codex API, and provides a completion back to the user presented in faded grey text which they may accept or reject (see \autoref{fig:assistant-ex}).

The general flow for using a coding assistant like ours or Copilot is as follows. 
The user types any amount of code, such as comments, function names and arguments, or implementations.
On the user pausing (750~ms of inactivity) the extension will select all code prior to their cursor and select text in reverse up to a finite amount---ours took up to 1,800 tokens (see \autoref{sec:background:llms}). It passes this text to the OpenAI Codex API for the \texttt{code-cushman-001} LLM. We chose this LLM as it is the fastest to operate and gave us response times similar to GitHub Copilot. To prompt the LLM we used the following parameters.
The \texttt{max\_tokens}: 64. This was chosen to keep the speed of generation high and the suggestions relatively short (we did not want the LLM to suggest code beyond the function currently being developed).
The \texttt{temperature}: 0.6. \newt{We set the temperature to 0.6 based on the results reported in the original Codex work~\cite{chen_evaluating_2021} as well as in \cite{doderlein_piloting_2022}, which show that the best pass@10 rate is at temperature=0.6. In addition,} this somewhat high temperature ensures the LLM does not provide the same answers to all users.
This is important as the same starting \texttt{list.c} file is provided to all users, and as our focus is on user acceptance of code suggestions rather than the LLM, it is beneficial if some suggestions by the LLM are unusual or creative. 
The \texttt{top\_p}: 1.0. OpenAI documentation suggests varying \texttt{temperature} or \texttt{top\_p}, not both.

Input (and settings) thus provided, the assistant responds with a code suggestion presented in gray italics (e.g. \autoref{fig:assistant-ex}). The user can accept the suggestion by pressing space bar or reject by continuing to type.

\subsection{`Autopilot'- automated task completion}
In addition to the two user groups, we created 30 solutions that were generated entirely by the Codex LLM as an `autopilot' group. 
We produced ten solutions from each of the three code LLMs offered by OpenAI: \texttt{code-cushman-001} (max 2048 tokens), \texttt{code-davinci-001} (max 4096 tokens) and \texttt{code-davinci-002} (max 8000 tokens). The last LLM is capable of filling in the middle given a prefix and suffix~\cite{bavarian_efficient_2022}). 

We queried the LLM to generate code for one function at a time in the order they appear in the template \texttt{list.c} file, requesting 512 tokens with a stop sequence of ``\verb|\n}\n|''. The prompt included the function declaration and as much of the previous file context as would fit in the LLM's context window (minus 512 tokens to allow room for the generated response), \emph{including} any code previously generated by the LLM. The \texttt{temperature} and \texttt{top\_p} were set to the same values as in the AI assistant IDE plugin (0.6 and 1.0) \newt{for the reasons indicated in the previous section}. 
After generating a function, we check if the result compiled. If compilation failed, we request another completion, up to a maximum of 10 attempts per function. If no code compiled, we used the template's implementation of the function, which just returns \texttt{EXIT\_FAILURE}.
This procedure models a user that relies on the AI assistant, accepting suggestions unconditionally, with minimal checks to see if the code compiles. Moving on to the next function once it seems to work and giving up if it fails after several attempts. This is our  baseline to compare control and AI-Assistant groups.
The initial file was identical to the starting template with one exception: we added a comment near the top of the file that listed the members of the \texttt{node} structure, which is defined in a header and is otherwise not visible to the LLM, as noted here:

\begin{lstlisting}[language=c]
// Members of the node struct:
// char* item_name, float price, int quantity, node *next
\end{lstlisting}

Without this addition, an unassisted LLM  must guess the member names, creating unusable solutions. This intervention is  realistic since our goal is to mimick a hands-off human user. Two users in the `assisted' group independently deployed this strategy by copying the \texttt{struct} definition from the header file into \texttt{list.c} (commented out).

We will present results for Autopilot group where appropriate, except for manual security analysis (\autoref{sec:results:subsec:security}), which was too time-consuming to expand to full set of 30 Autopilot solutions. Instead, we audited five Cushman LLMs.

\subsection{Experimental infrastructure}
\newt{This study provided a controlled, consistent environment via a containerized cloud-based IDE usable inside standard web browsers, based on the open-source Anubis software\footnote{https://github.com/AnubisLMS/Anubis} which is commonly used at New York University. %
This environment contained a virtual Visual Studio Code instance, which automatically loaded the project upon a user signing in.
Based on their group membership, the system would automatically either provide code suggestions or not. %
This custom IDE made it straightforward to add data collection for active participants and prevent the IDE from being connected to other LLMs.}

In addition to the `final form upload', which collected the self-reported `finished' \texttt{list.c} files (as well as collected the study's demographic information), we also took snapshots of the complete \texttt{list.c} file environment every 60 seconds that the file was open. This allowed us to track changes over time.
In addition, we recorded when suggestions from the Codex-based AI assistant were taken or when they were rejected.

\subsection{Statistical tests}
We use standard statistical hypothesis tests to analyze results. We check if LLM code assistants improve code quality without exacerbating security. Analogously, in medical settings, it is required to show that a treatment is effective (code quality) while not exacerbating side effects (security). Standard comparative tests are used to establish efficacy--- i.e., that the mean efficacy of the treatment is higher than the control.
For side effects, non-inferiority tests are used---the test seeks to establish that the side effects are within a ``maximum clinically acceptable difference" that one is willing to tolerate~\cite{walker_understanding_2011}.

We describe the two tests below.

\textbf{Comparative hypothesis test:} 
Given treatment and control groups with means $\mu_{1}$ and $\mu_{0}$, respectively, and assuming (without loss of generality) that smaller means are better, a comparative hypothesis test seeks to reject the null hypothesis $H_{0}$ in favor of the alternate $H_{1}$. 
\begin{itemize}
    \item Null hypothesis ($H_{0}$): Treatment and control groups have the same mean, i.e., $\mu_{1}=\mu_{0}$.
    \item Alternate ($H_{1}$): Treatment group has a lower mean that control, i.e., $\mu_{1} < \mu_{0}$.
\end{itemize}
A comparative test establishes that the treatment is ``better" than the control. We use this test to compare assisted (i.e.,  treatment) and control groups in terms of number of compiling functions and unit tests passed (in both cases, larger is better, so the hypothesis test is modified accordingly.).

\textbf{Non-inferiority hypothesis test:} 
Under the same assumptions as above, the null and alternate hypotheses of a non-inferiority test are:
\begin{itemize}
    \item Null hypothesis ($H_{0}$): The treatment mean is more than $\delta \%$ larger than control, i.e., $\mu_{1} > (1+\frac{\delta}{100}) \times \mu_{0}$,
    \item Alternate ($H_{1}$): The mean of the treatment group is less than $\delta \%$ larger than control, i.e., $\mu_{1} < (1+\frac{\delta}{100}) \times \mu_{0}$,
\end{itemize}
 where $\delta$ is the tolerance threshold. We use non-inferiority tests to compare bug incidence, measured as CWEs/LoC (\autoref{eq:cwe-per-loc}, \autoref{sec:result-metrics}) and average CWEs/function (\autoref{eq:cwe-per-n}), in the assisted and control groups. 
 
 There is no commonly accepted threshold for the amount of decrease in code security that is considered acceptable for a new programming tool. Different organizations may make different choices depending on their threat model. 
For this study we pick a threshold of 10\% (i.e., we test the hypothesis that AI assistance introduces no more than 10\% more vulnerabilities per line of code), but we make our data and analysis available so that other thresholds can be tested if desired. \newt{The 10\% threshold was chosen due to its common use in prior work (e.g. \cite{hung_regulatory_2005} in medical studies). As  \cite{hung_regulatory_2005} notes, ``This margin has to incorporate a value judgment that...can only be made subjectively and involves benefit/risk assessments...In the medical community, this margin has been reached via consensus.'' We hope to see a similar effort in the software security community to establish acceptable consensus values.} With $\delta=10\%$,   
 rejecting the null confirms that the bug incidence in the Codex assisted group is at worst $10\%$ greater than the control group; we believe that this would show that Codex assistance does not exacerbate security ``too much.''

%% file: sections/04Results.tex
\section{Study results and analysis}
\label{sec:results}

\subsection{User population (Demographics)}
\label{sec:demographics}

As noted (\autoref{sec:recruitment}), 58 participants completed code for analysis.
As part of the study the participants completed a brief demographic questionnaire.
\autoref{tab:participant-demographics} shows the academic enrolment information broken down by study group. 
There was a good balance between undergraduates (UG), postgraduates (PG), and others (e.g. recent graduates) across the two study groups (`control' and `assisted').

\begin{table}[]
\caption{Study participant enrolment demographics}
\label{tab:participant-demographics}
\resizebox{1\linewidth}{!}{%
\input{tables/demographics}}
\end{table}

To examine pre-existing participant knowledge, we asked the three questions presented in \autoref{tab:participant-demographics-experience}.
The first question checks if the assignment was similar to previous work completed by the participants, and about half of each of the `control' and `assisted' groups self-reported having written a linked list in C before.
The latter two questions aimed to validate our goals with participant recruitment (i.e., they should have some experience with C and knowledge of the linked list data structure). The majority of participants in both study groups had both written C code before and had previously or were currently taking a data structures or algorithms class.

\begin{table}[]
\caption{Study participant experience demographics}
\label{tab:participant-demographics-experience}
\resizebox{1\linewidth}{!}{%
\input{tables/experience-demos}}
\end{table}

\subsection{RQ1 - Functionality}
\begin{figure}[b]
\centering
\includegraphics[width=\linewidth]{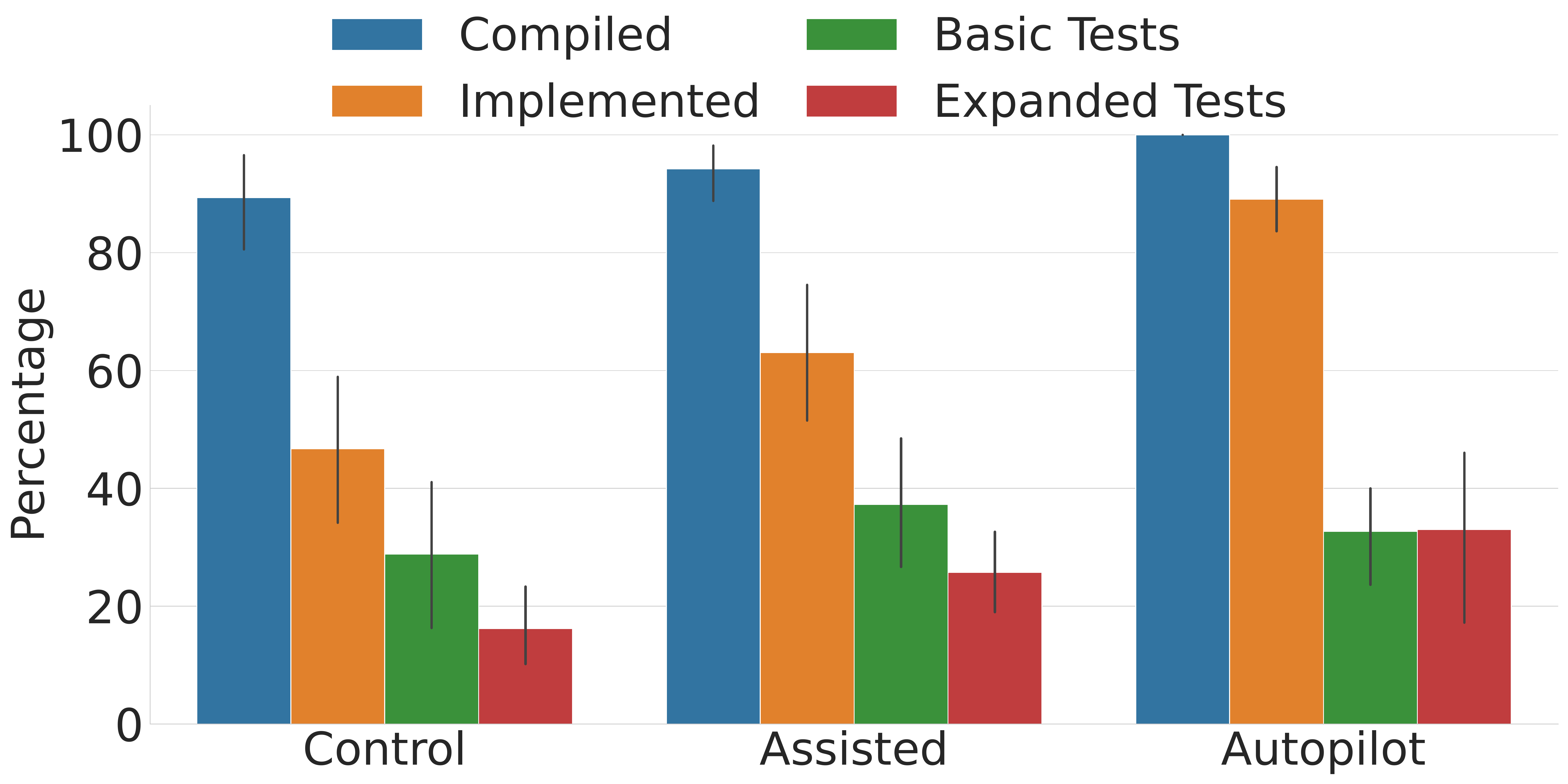}
  \caption{Functionality for each group. Each group has to implement 11 functions and 11 basic tests. We had 43 expanded tests. We show per group, the average \% of functions \textbf{Implemented}, regardless of whether they compiled or not, the average \% of those functions that \textbf{Compiled}, the average \% of each group that passes the 11 \textbf{Basic Tests} and the average \% that passes the 43 \textbf{Expanded Tests} from each group. }
\label{fig:functionality}
\end{figure}

We assess functionality of the code using unit tests. Besides the 11 \textbf{Basic Tests} (see \autoref{fig:functionality})
that  we provided to the participants,  we wrote 43  \textbf{Expanded Tests} to exercise edge cases (e.g., adding an element to the head of the list, providing the same position for both arguments to \texttt{list\_swap\_ item\_positions}), invalid parameters (e.g., NULL pointers, zero/negative indices), and validating that return value and state of the list are correct after each API call.

\textbf{Split Testing:} We faced two challenges in automatically testing the functionality of submitted code. First, many submissions did not compile (19/58 = 32.8\%). This includes  9/30=30\% in the Assistant group and 10/28=35.7\% in the Control group). Second, the tests in the test suite may need to use other API functions in addition to the one under test (e.g. a test for \texttt{list\_delete\_item\_at\_pos} might need to create a nonempty list by using \texttt{list\_init} and \texttt{list\_add\_item\_at\_pos}). If there is a serious bug in one of the core functions, it will cause all tests that depend on it to fail (even if the function under test itself is correct). It is difficult to measure functionality of the program as a whole.

Our testing procedure solves both of these problems by splitting the user's code into individual functions (one per API, including any required helper functions or data structures). For each function under test, we create a version of the submitted code where the other API functions are replaced by our own known-good reference implementations. %
If an extracted function does not compile, we mark it as non-compiling and mark its associated tests as failing. 
As the template code ``implements'' each function by returning \texttt{EXIT\_FAILURE}, and some tests expect failure, unmodified code may spuriously pass some tests. So, we also require code modification.

Our tests thus \emph{automatically} measure four distinct quantitative aspects of functionality for each submission:
(i) \% functions that were implemented,
(ii) \% functions that compiled,
(iii) \% basic tests passing, and
(iv) \% expanded tests passing.
The four measures are shown for each group (`control', `assisted', and `autopilot') in \autoref{fig:functionality}.

\textbf{Results:} We see systematic differences between the `assisted' and the `control' group. 
the `assisted' group had a small but consistent advantage over the `control' group. The 'autopilot' group outperformed both `assisted' and 'control' groups on functions implemented and compiled---this is by design since our autopilot code generation procedure repeats several times till code compiles. Interestingly, `autopilot' slightly under performs `assisted' on basic tests, but slightly over performs on expanded tests. 
In other words, the AI code assistant does help the users write better code in terms of functionality.

Finally, the `assisted' group wrote more code overall (280.9 average lines of code compared to 247.5 LoC in the control group). We note, however, that due to the small sample size none of these comparisons reach statistical significance at the standard $p<0.05$ level (as tested using Fisher's exact test for the completion data, which is a binary variable, and Welch's t-test for the remaining comparisons). 

\newt{For interest, we include a graph indicating the time at which users in the different groups completed the study, with the assisted group writing their submissions faster. This is depicted in \autoref{fig:study-duration}, in the Appendix.}

\subsection{RQ2 - Security analysis}
\label{sec:results:subsec:security}

Although there are many different tools available for finding security-relevant flaws in C source code (as discussed in \autoref{sec:background:subsec:codesec}), we found that none of them were appropriate for our use case. 
Static analysis tools such as CodeQL~\cite{github_inc_codeql_2021} gave rates of false positives and negatives too high for our purposes. 
Meanwhile, fuzzing the participant code created records difficult to deduplicate (this is an open research problem~\cite{manes_art_2021}). Further, as fuzzing is dynamic, any vulnerability causing a crash along a program path rendered vulnerabilities later in the path unreachable, underestimating the true vulnerability count. 
For these reasons, we opted to manually audit the 58 user-generated submissions, and five of the \texttt{code-cushman-001} LLM answers for comparison, a process further discussed here.

\subsubsection{Bug data encoding}
\label{sec:bug-id-method}

Working one function at a time, a panel of three of the co-authors collectively read through blinded copies of submitted source code, annotating security-relevant bugs as comments.
This process was guided by compiler logs and the basic and extended test suites; as well as thorough manual lexical analysis. 
\newt{As all annotation was performed collectively, no inter-rater reliability checks needed to occur}.
This manual audit took about 22 hours over the course of one week translating into 66 person-hours overall---manual analysis, while thorough, does not scale well.

From this, we created a table of 67 unique bug classes across all functions. Our focus was on memory related or undefined bugs that cause CWEs such as the ones on \autoref{table:cwe-list}. %
Full records with all annotations are provided open source~\cite{sandoval_lost_2022}.
A summary, with severe bugs found per function with incidence rates is in \autoref{tbl:severe-cwe-counts-per-function} (in the Appendix).

\textbf{Example bug finding process}: One study participant provided the code exactly as it appears in  \autoref{fig:assistant-ex} for \texttt{list\_item\_to\_string}. 
This was a second-year UG student who had written C code before and took an algorithms class. They were in the `assisted' group.

This code passes basic functional tests.
However, it has three CWEs.
The first weakness is CWE-476: \textit{NULL Pointer Dereference}. This can occur in the case where \texttt{str} is NULL when this function is called (this is not checked for in the code, and the API cannot guarantee the values it will be passed as arguments).  This CWE is ranked at position \#11 on Mitre's 2022 `Top 25' list~\cite{mitre_2022_2022}.
The next weakness is CWE-758: \textit{Reliance on Undefined, Unspecified, or Implementation-Defined Behavior}. This can occur when \texttt{head->item\_name} is NULL, and occurs because \texttt{sprintf} does not define what should happen when NULL is passed to the \texttt{`\%s'} argument. This is a minor (some would say negligible) issue, as in the standard libraries for \texttt{gcc} \texttt{sprintf} will print it as \texttt{(null)}.

The third and final weakness is the most serious. It is CWE-787: \textit{Out-of-bounds Write}, ranked as \#1 on Mitre's `Top 25' list.
This occurs because the function \texttt{sprint}s to an \emph{externally allocated} string. 
What is the length of this string? 
It is defined in a \texttt{\#define} at the top of \texttt{list.c} (see \autoref{fig:lst:hints}, line 7).
This is important because \texttt{head->item\_name} is a \emph{user-controlled} value, meaning they could store a very long string in here which would run off the end of the buffer.
The only safe way to implement this function is to use \texttt{s\textunderscore{n}printf} with the $n$ set to the value of this \texttt{\#define}.
This function is scored as \texttt{passing} the basic tests, \texttt{failing} extended tests (they pass in NULL as \texttt{str} for one case), and features \emph{three} CWEs, \emph{two} ranked as severe.

\subsubsection{Metrics}
\label{sec:result-metrics}
We analyze the quality of each user's submission using the bug per line-of-code (LoC) as our metric. Since we associate each bug with a CWE, we will use CWEs/LoC as the metric. Since most users submitted valid implementations for a subset of the 11 functions they were tasked with implementing, we compute CWEs/LoC over those functions. We adopt two notions of validity: (1) the function compiles or (2) it compiles \emph{and} passes unit tests.

The CWEs/LoC are computed as follows. if $E_{ij}$ is the number of CWEs in function $j$ of user $i$'s submission,  $V_{ij} \in \{0,1\}$ is a binary variable that is one only if user $i$ submitted valid code for function $j$, and $L_{ij}$ are the LoCs written by user $i$ for function $j$, then the \textbf{CWEs/LoC} for user $i$ are:
\begin{equation}
\label{eq:cwe-per-loc}
    M^{1}_{i} = \frac{ \sum_{j=1}^{11} {E_{ij} V_{ij}} } {   \sum_{j=1}^{11} {L_{ij} V_{ij}} }.
\end{equation}
We report {CWEs/LoC} in 'assisted' and 'control' groups by averaging $M^{1}_{i}$ over users in these groups. 
We compute \textbf{Severe CWEs/LoC} metric focusing on the top-25 security CWEs reported by Mitre~\cite{mitre_cwe_2022}. The results are shown in Figure \ref{fig:aggregate}.

Since our methodology allows us to test each function independently, we can compare CWE incidence rates on a \emph{per function} basis. For this, we compute the \textbf{average CWEs for function $j$} as:
\begin{equation}
\label{eq:cwe-per-n}
  M^{2}_{j}(g) = \frac{\sum_{ i \in {N_{g}}} E_{ij} V_{ij}}{ \sum_{ i \in {N_{g}}} V_{ij}}, \quad g \in \{\texttt{Assist}, \texttt{Control}\} 
\end{equation}
where $N_{\texttt{Assist}}$ and $N_{\texttt{Control}}$ are Codex assisted and control group users, respectively.

\subsubsection{Topline results---CWEs/LoC}

\begin{figure}[tbp]
  \begin{subfigure}[t]{0.22\textwidth}
    \centering
    \includegraphics[width=\linewidth]{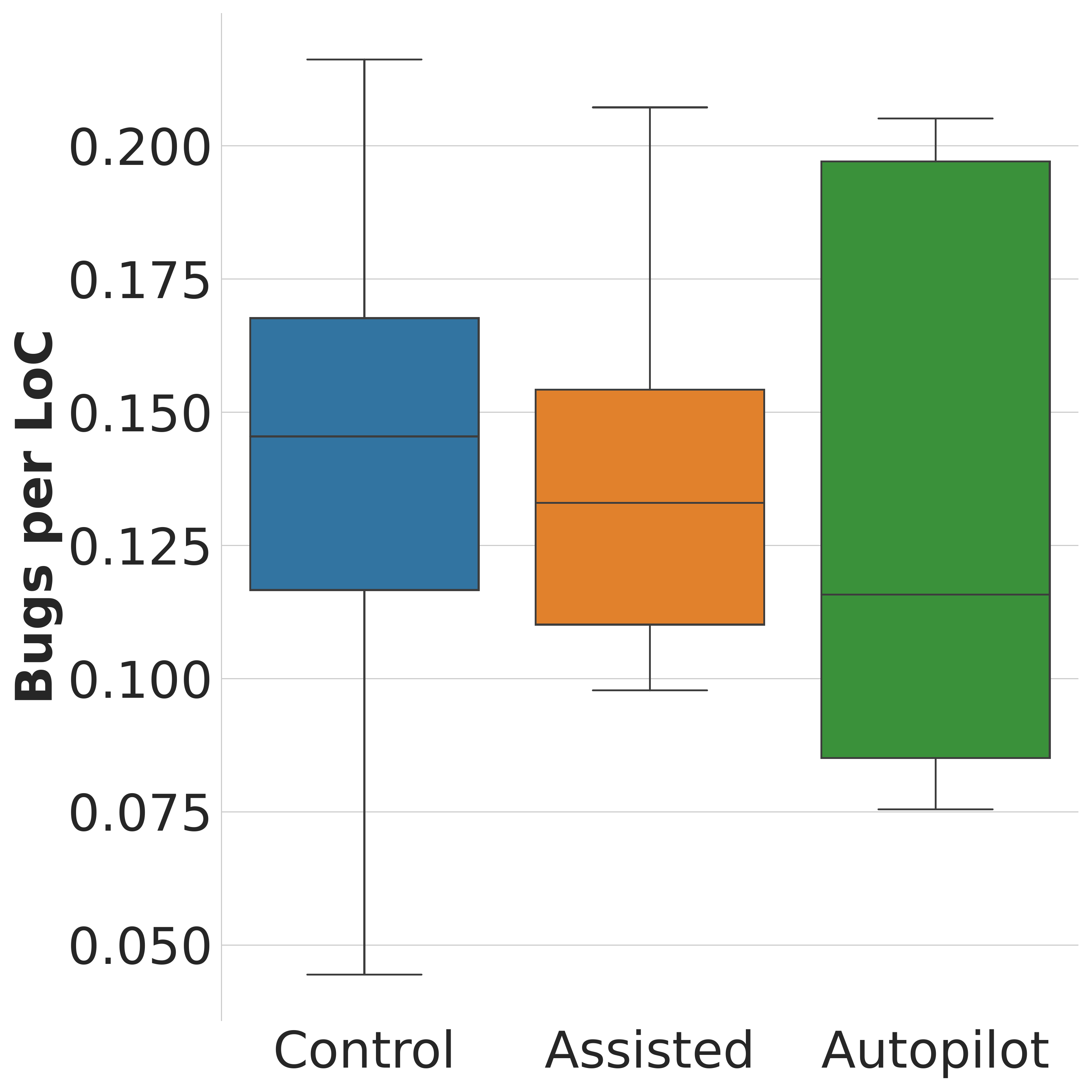}
    \caption{\textbf{CWEs/LoC} over compiling functions. \label{fig:aggregate-a}}
  \end{subfigure}
  \hspace*{\fill}
  \begin{subfigure}[t]{0.22\textwidth}
    \centering
    \includegraphics[width=\linewidth]{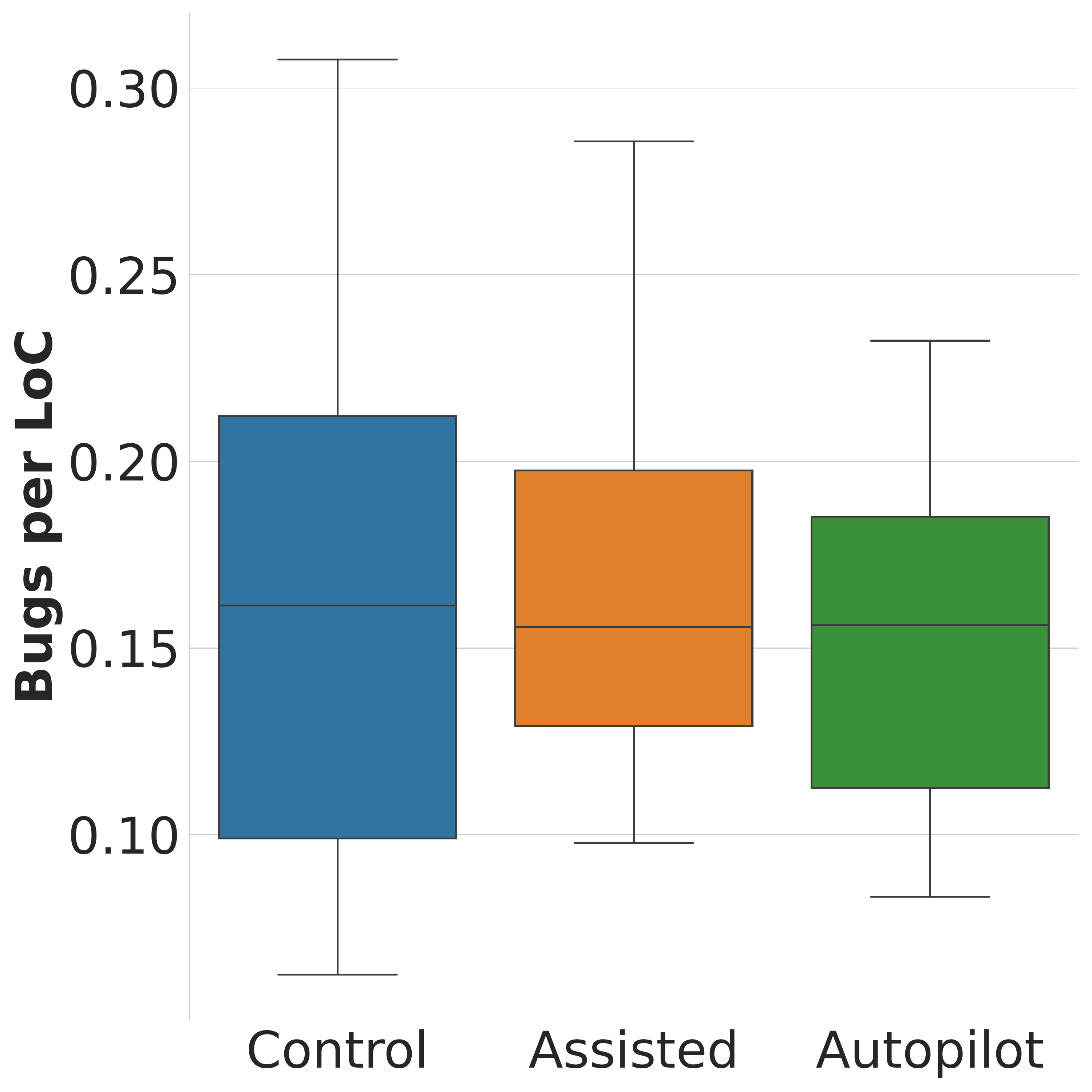}
    \caption{\textbf{CWEs/LoC} over functions that pass unit test. \label{fig:aggregate-b}}
  \end{subfigure}
  \hspace*{\fill}
  \\
  \begin{subfigure}[t]{0.22\textwidth}
    \centering
    \includegraphics[width=\linewidth]{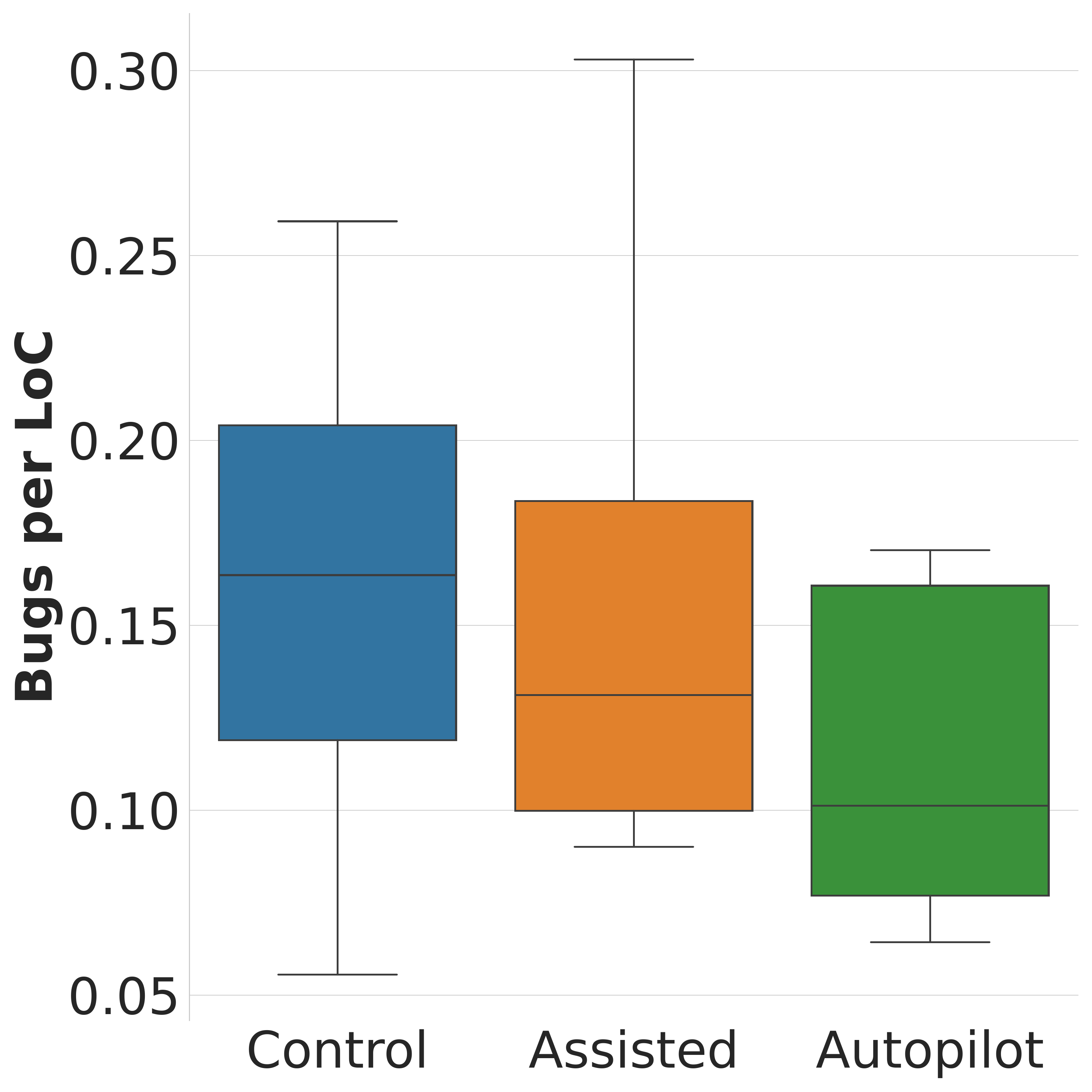}
    \caption{\textbf{Severe CWEs/LoC} over compiling functions. Non-inferiority test is significant with $p=0.04$. \label{fig:aggregate-c}}
  \end{subfigure}
  \hspace*{\fill}
  \begin{subfigure}[t]{.22\textwidth}
    \centering
    \includegraphics[width=\linewidth]{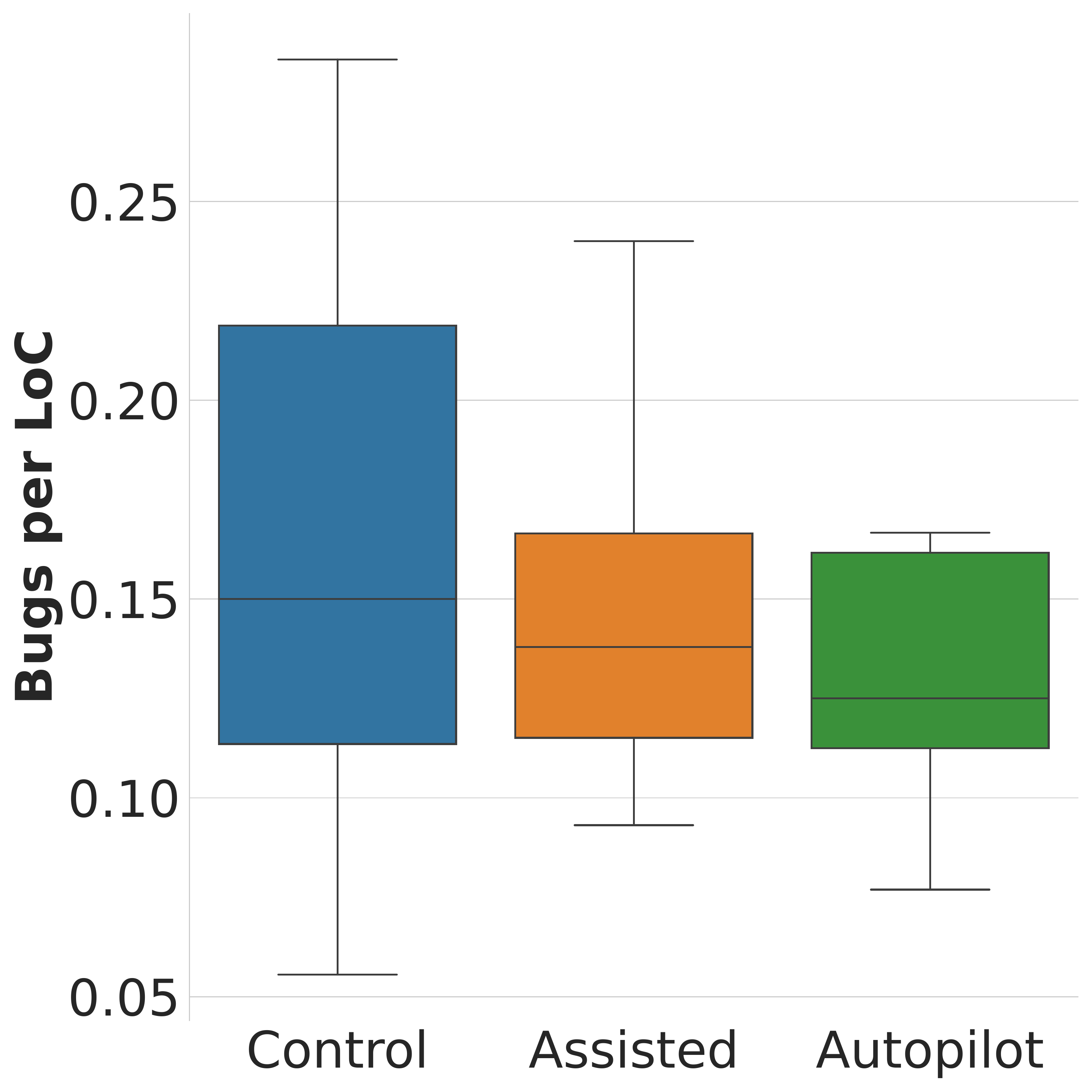}
    \caption{\textbf{Severe CWEs/LoC}  over functions that pass unit test.  \newt{Non-inferiority test gives $p=0.06$.}} 
    \label{fig:aggregate-passing-severe-per-loc}
  \end{subfigure}
  \medskip
\caption{Comparing CWEs/LoC over compiling/passing functions for `all CWEs' and `severe CWEs' for each group. 
Statistically significant p-values between `assisted' and `control' for non-inferiority test ($\delta=10\%$) are noted.
}
\label{fig:aggregate}
\end{figure}

\autoref{fig:aggregate-a}-\autoref{fig:aggregate-b} shows boxplots of the CWEs/LoC over compiling functions and functions passing unit tests for the three groups, while \autoref{fig:aggregate-c}-\autoref{fig:aggregate-passing-severe-per-loc} do the same for severe CWEs.
For all four cases, we found that the 'assisted' group has fewer bugs compared to 'control', with up to a 22\% lower mean for the 'assisted' group compared with the 'control' for severe CWEs over passing tests.
For severe CWEs, the comparisons are also statistically significant using non-inferiority tests with $\delta=10\%$, i.e., we can conclude that severe bugs/LoC for the 'Assisted' group are no more than $10\%$ greater than in the 'Control' group.

\subsubsection{Per function CWE rates}
Our topline results suggest that CWE incidence in `assisted' and `control' groups are close. We now check whether these groups differ at the function level, i.e., whether Codex assisted users introduce more bugs in certain functions compared to the control group or vice-versa. 
The per-function CWE rates (see \autoref{eq:cwe-per-n} to see how compute this), %
of severe CWEs (those within Mitre's `Top 25') written by the three study groups are presented in \autoref{tbl:results-cwe-counting}.
We present the results for (a) all functions that compile, and (b) the compiling functions that then go on to pass the basic suite of tests.
We present the number of passing functions from each group, as well as the number of CWEs found in those functions. The rate is the division of these two numbers (see \autoref{eq:cwe-per-n}).
Security-related errors within functions that pass tests are naturally more concerning than those in functionally-buggy code, as code that appears to be functional has a higher chance of being deployed as-is.

As shown, the results for individual function vary across groups. Where the `assisted'
group has a $10\%$ higher rate of bugs compared to the `control' is highlighted in blue. Over functions that pass unit tests, `assisted' users have more bugs for functions that perform input/output operations on the linked list. %
Conversely, functions for which the `control' group has $10\%$ higher rate of bugs compared to `assisted' are highlighted in blue. %
These functions tend to involve pointer manipulations and/or more complex logic.
\newt{Where a difference between the two groups is statistically significant, we annotate the fields with a $\dagger$.}

In terms of absolute CWE rates, for the `control' and `assisted' groups, \texttt{list\_add\_ item\_at\_pos}, \texttt{list\_remove\_item\_at\_pos}, and \texttt{list\_update\_\allowbreak item\_at\_pos} feature significantly higher incidence rates of severe CWEs ($\approx$50\,\% greater than the `average' function rate), likely reflecting the increased difficulty when writing pointer and string manipulations (notoriously fiddly in C).
For interest, we also include \autoref{tbl:severe-cwe-counts-per-function} (in the Appendix) which presents the number of each type of CWE identified in the different functions by each study group.

\begin{table}[h]
\caption{Counting the \# of Severe CWEs identified in each function/group. N = (N)umber of submitted functions which compile / which pass the tests. Rate is the average of the Severe CWEs count per function in this group. Yellow cells indicate `control' has 10\,\% higher rate of bugs than `assisted', Blue is reverse. \newt{Cells with $\dagger$ indicate where this difference is statistically significant}. The `Autopilot' group in this Table describes only the first 5 \texttt{code-cushman-001} answers.}
  \centering
  \label{tbl:results-cwe-counting}
\resizebox{\linewidth}{!}{\input{tables/cwes}}
\vspace{-6mm}
\end{table}

\subsubsection{CWE incidence rates}
\label{sec:cwe-incidence}
\begin{figure}
\centering
  \includegraphics[width=\linewidth]{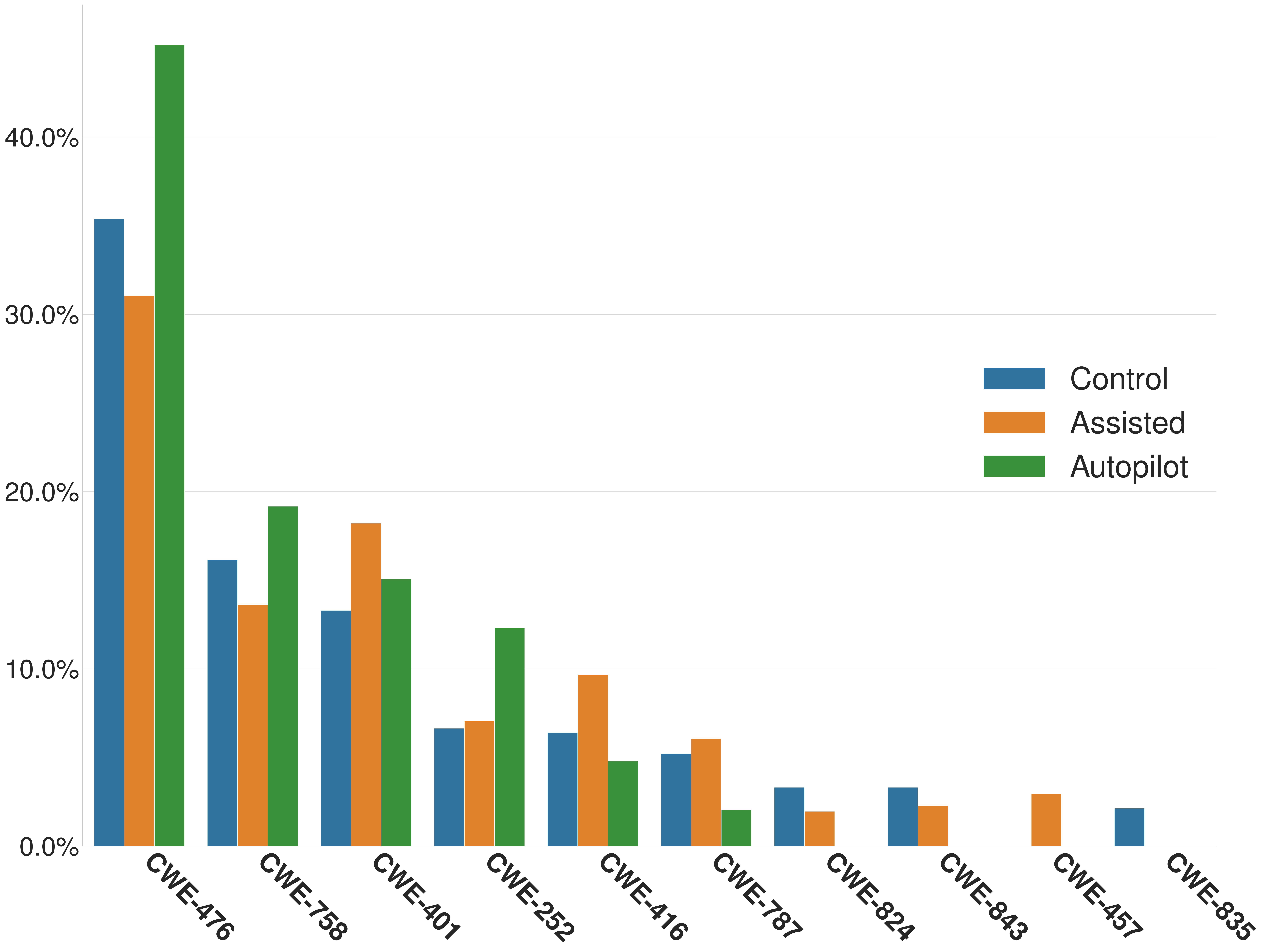}
  \caption{Top 10 CWEs per group and their prevalence as \% of total CWEs. CWE descriptions are in \autoref{table:cwe-list}.}
\label{fig:cwe_prevalence}
\vspace{-4mm}
\end{figure}

\autoref{fig:cwe_prevalence} shows the prevalence of the ten most common CWEs in user submissions. %
CWE descriptions in \autoref{table:cwe-list} in the Appendix, along with their severity rank---if a CWE is not first-order severe, a possible second-order severity is presented alongside.
For instance, with CWE-401, which is unranked, the downstream effect is CWE-400, ranked at (\#23).

CWE-787 (out-of-bounds write), the most severe CWE, is about equally prevalent in the `control' and `assisted' groups, but far less prevalent in the `autopilot' group. This is likely due to the main root cause of CWE-787, which was, by far, most frequently caused by the use of \texttt{sprintf} rather than \texttt{snprintf} (e.g. in in \texttt{list\_item\_to\_string} as previously discussed in \autoref{sec:bug-id-method}). 

CWE-416 (use after free), the second most severe CWE, is more prevalent in the `assisted' group compared to the `control'. This appears to be due to mistakes frequently made by the `assisted' group when manipulating the handling of the \texttt{char* item\_name} fields.
Here, when creating a new shopping list node, the \texttt{item\_name} is passed as an argument.
There are many ways that this name could be stored.
The unsafe and na\"{i}ve way is to copy the \texttt{char*} pointer values.
As the API is not provided any guarantees about the memory location this is pointing to, it is not safe to assume that this value will persist beyond the call of this function.
Performing a pointer copy may thus lead to CWE-416 if the memory is later freed (there will now be a dangling pointer to the freed memory).
The `safe' way to manage the \texttt{item\_name} variable is to perform a \emph{copy} of the string into a new variable.
There are two reasonable methods---the first (and easiest) would use \texttt{strdup}, and the second would use \texttt{strlen} followed by a \texttt{malloc(strlen+1)} followed by a \texttt{strcpy}.

Across all submissions, CWE-476 (NULL Pointer Dereference) was the most commonly observed potential vulnerability.
This is because the API of the shopping list does not guarantee argument correctness: so every single pointer should always be checked against NULL.
Special cases are where a function takes a double-pointer, such as \texttt{list\_add\_item\_at\_pos} taking \texttt{node** head}. Here, both \texttt{head} and \texttt{head*} need to be checked against NULL. Such requirements were often missed by all participants, human and LLM. These had downstream effects. For example, it causes a large proportion of the CWE-758 instances (Reliance on Undefined Behavior), a CWE frequently observed when code uses standard library functions that may ingest NULL pointers (e.g. \texttt{printf, strcpy, strlen}).

\subsubsection{Observations}

The impact of code suggestions on cybersecurity (RQ2) is less conclusive than the impact on functionality (RQ1). 
\autoref{tbl:results-cwe-counting} suggest that certain kinds of functions may be more or less difficult to write safely depending on their complexity and the experience of the developer---it appears that the LLMs may sometimes reduce the incidence rates of bugs, and sometimes increase them.
Meanwhile, aggregating CWEs per participant LoC (\autoref{fig:aggregate})  suggests that there may be a slight benefit to using LLMs, with \autoref{fig:aggregate-passing-severe-per-loc} in particular highlighting that as code is made to pass tests it may be made more secure by using the LLM.
This is contrary to literature~\cite{pearce_asleep_2022,siddiq_empirical_2022} which suggests that LLMs should be used with care due to their habit of suggesting vulnerable patterns.

\subsection{RQ3 - On the origin of bugs}

To better understand how LLM assistance contributed to the code written by users in the Assisted group, we created a visualization tool (\autoref{fig:suggestion_viz} in the Appendix) that colors each user's code based on the accepted suggestions logged during the experiment (ignoring code from our provided initial template). The tool takes each suggestion in reverse chronological order (most recent first) and attempts to match it to a portion of the final document, initially using exact string matching and then attempting approximate matches up to a normalized edit distance of 50\%. With code originating from the LLM, the original suggestion will be shown on hover.

Using this tool, we examined each of the 564 security vulnerabilities identified in \autoref{sec:cwe-incidence} and coded them as `originating from a Codex suggestion' or as `introduced manually by a human user' (\autoref{tab:attrib}). We found that humans introduced 356 of the bugs in our dataset (63\%), while only 36\% were introduced by the LLM and present in the user's code verbatim (16\%) or with modifications (20\%).

Overall, 60\% of the non-template code was written by a human. This accords with our findings in \autoref{sec:cwe-incidence}: the rate at which vulnerabilities are introduced by the LLM is similar to the rate at which they are introduced by humans. This makes sense intuitively: LLMs attempt to predict the most likely continuation of their input, and so the quality of the code they output tends to match the quality of their input.

\begin{table}[]
\centering
\caption{Attribution of bugs and LoCs for Codex-assisted users. Approx. is code manually modified from a suggestion.\\}
\label{tab:attrib}
\vspace{-2mm}
\resizebox{0.55\linewidth}{!}{%
\begin{tabular}{|l|r|r|r|}
\hline
    & \multicolumn{1}{l|}{\textbf{Human}} & \multicolumn{1}{l|}{\textbf{Codex}} & \multicolumn{1}{l|}{\textbf{Approx.}}  \\ \hline
\textbf{LoC} & \textbf{60\% }                               & \textbf{16\% }                               & \textbf{18\%}  \\

\textbf{Bugs} & \textbf{63\% }                               & \textbf{16\% }                               & \textbf{20\%}                                                                 \\ \hline
\end{tabular}
}
\vspace{-5mm}
\end{table}

We can qualitatively examine the origin of a single bug. As an example, consider the potential use-after-free CWE-416 discussed in \autoref{sec:cwe-incidence}.
\newt{This is chosen as incidences of this bug were lexically similar between users and suggestions.}%
We choose this to analyze as it is a bug which is straightforward to test for using a custom automated script. %
How do users interact with buggy suggestions from the LLMs, and how bugs might `amplify' via LLM suggestions if  present in code.

\newt{To examine this, we first identify the CWE-416 incidences using the annotated final files from RQ2. We then progressively can scan the document and suggestion snapshots recorded during the user study, looking for the first recorded incident of that bug---for example,} the first time that a new node's item name is incorrectly set directly to the function argument \texttt{item\_name} (i.e. without using a proper string copy mechanism). 
We then count the number of times that the bug was present in suggestions by the LLM, as well as the number of suggestions containing the bug that were accepted by the user.
As user acceptance of suggestions is still not fully reflective of the final state of the code (as accepted code may be further edited), we also scan the final `finished' code files to count the number of these bugs present.
Note that this bug can occur in multiple locations---both \texttt{list\_add\_item\_at\_pos} and \texttt{list\_update\_item\_at\_pos} need to copy item names. %

\begin{table}[t]
\caption{Origins of CWE-416 `use after free' bug when \texttt{item\_name} is improperly copied by users in `assisted' group.}
  \centering
  \label{tbl:results-bug-origins}
\resizebox{0.85\linewidth}{!}{\input{tables/bug-origins}}
\vspace{-3mm}
\end{table}

We report results of this investigation in \autoref{tbl:results-bug-origins}. Looking at this bug, in most cases it comes from the LLM suggestion originally, and even when it appears in the document first, the LLM will go on to suggest the bug.
Users that had the highest number of this bug had the highest number of buggy suggestions provided and also accepted the highest number of suggestions. This table provides some insights into the usage of the LLM in general: users `1f1c`, `2125', `a5ba', and `dc47' self-author the bug without suggestions and do not go on to accept buggy suggestions from the LLM (nor in many cases even generate relevant suggestions).

%% file: tables/demographics.tex
\begin{tabular}{rccl}
\cline{2-4}
\multicolumn{1}{l|}{}                        & \multicolumn{1}{c|}{\textbf{Control}}    & \multicolumn{1}{c|}{\textbf{Assisted}}    & \multicolumn{1}{c|}{\textbf{Total}}                \\ \hline
\multicolumn{4}{|c|}{\textit{Undergraduates (UG)}}                                                                                                                                                   \\ \hline
\multicolumn{1}{|r|}{UG Y2 (Sophomores)} & \multicolumn{1}{c|}{\textit{2}}                & \multicolumn{1}{c|}{\textit{7}}                 & \multicolumn{1}{l|}{}                              \\ \hline
\multicolumn{1}{|r|}{UG Y3 (Juniors)}    & \multicolumn{1}{c|}{\textit{8}}                & \multicolumn{1}{c|}{\textit{5}}                 & \multicolumn{1}{l|}{}                              \\ \hline
\multicolumn{1}{|r|}{UG Y4 (Seniors)}    & \multicolumn{1}{c|}{\textit{4}}                & \multicolumn{1}{c|}{\textit{4}}                 & \multicolumn{1}{l|}{}                              \\ \hline
\multicolumn{1}{|r|}{UG (Unspecified)}       & \multicolumn{1}{c|}{\textit{2}}                & \multicolumn{1}{c|}{\textit{1}}                 & \multicolumn{1}{l|}{}                              \\ \hline
\multicolumn{1}{|r|}{\textbf{UG (Total)}}    & \multicolumn{1}{c|}{\textbf{16}}               & \multicolumn{1}{c|}{\textbf{17}}                & \multicolumn{1}{l|}{\textbf{N (UG) = 33}}          \\ \hline
\multicolumn{4}{|c|}{\textit{Postgraduates (PG)}}                                                                                                                                                    \\ \hline
\multicolumn{1}{|r|}{PG (MS)}           & \multicolumn{1}{c|}{\textit{10}}               & \multicolumn{1}{c|}{\textit{10}}                & \multicolumn{1}{l|}{}                              \\ \hline
\multicolumn{1}{|r|}{PG (PhDs)}              & \multicolumn{1}{c|}{\textit{1}}                & \multicolumn{1}{c|}{\textit{1}}                 & \multicolumn{1}{l|}{}                              \\ \hline
\multicolumn{1}{|r|}{PG (Unspecified)}       & \multicolumn{1}{c|}{\textit{1}}                & \multicolumn{1}{c|}{\textit{0}}                 & \multicolumn{1}{l|}{}                              \\ \hline
\multicolumn{1}{|r|}{\textbf{PG (Total)}}    & \multicolumn{1}{c|}{\textbf{12}}               & \multicolumn{1}{c|}{\textbf{11}}                & \multicolumn{1}{l|}{\textbf{N (PG) = 23}}          \\ \hline
\multicolumn{4}{|c|}{\textit{Other Participants}}                                                                                                                                                    \\ \hline
\multicolumn{1}{|r|}{\textbf{Other (Total)}} & \multicolumn{1}{c|}{\textbf{1}}                & \multicolumn{1}{c|}{\textbf{1}}                 & \multicolumn{1}{l|}{\textbf{N (Other) = 2}}        \\ \hline
\multicolumn{1}{|r|}{\textbf{Total}}         & \multicolumn{1}{l|}{\textbf{N (Control) = 29}} & \multicolumn{1}{l|}{\textbf{N (Assisted) = 29}} & \multicolumn{1}{l|}{{\ul \textbf{N (Total) = 58}}} \\ \hline
\multicolumn{1}{l}{}                         & \multicolumn{1}{l}{}                           & \multicolumn{1}{l}{}                            &                                                   
\end{tabular}

%% file: tables/experience-demos.tex
\begin{tabular}{rccc|}
\cline{2-4}
\multicolumn{1}{l|}{}                              & \multicolumn{1}{c|}{\textbf{Control}} & \multicolumn{1}{c|}{\textbf{Assisted}} & \textbf{Total} \\ \hline
\multicolumn{4}{|c|}{\textit{Is this the first linked list implementation you have ever made in C?}}                                                             \\ \hline
\multicolumn{1}{|r|}{\textbf{Yes (first list)}}    & \multicolumn{1}{c|}{\textit{15}}            & \multicolumn{1}{c|}{\textit{14}}             & 29             \\ \hline
\multicolumn{1}{|r|}{\textbf{No (not first list)}} & \multicolumn{1}{c|}{\textit{11}}            & \multicolumn{1}{c|}{\textit{12}}             & 23             \\ \hline
\multicolumn{1}{|r|}{\textbf{Declined to answer}}  & \multicolumn{1}{c|}{\textit{3}}             & \multicolumn{1}{c|}{\textit{3}}              & 6              \\ \hline
\multicolumn{4}{|c|}{\textit{Is this the first time that you have ever programmed in C?}}                                                                        \\ \hline
\multicolumn{1}{|r|}{\textbf{Yes (first time)}}    & \multicolumn{1}{c|}{\textit{3}}             & \multicolumn{1}{c|}{\textit{4}}              & 7              \\ \hline
\multicolumn{1}{|r|}{\textbf{No (not first time)}} & \multicolumn{1}{c|}{\textit{23}}            & \multicolumn{1}{c|}{\textit{21}}             & 44             \\ \hline
\multicolumn{1}{|r|}{\textbf{Declined to answer}}  & \multicolumn{1}{c|}{\textit{3}}             & \multicolumn{1}{c|}{\textit{4}}              & 7              \\ \hline
\multicolumn{4}{|c|}{\textit{Are you taking, or have you ever taken a data structures or algo. class?}}                                          \\ \hline
\multicolumn{1}{|r|}{\textbf{Currently taking}}    & \multicolumn{1}{c|}{3}                      & \multicolumn{1}{c|}{2}                       & 5              \\ \hline
\multicolumn{1}{|r|}{\textbf{Previously taken}}    & \multicolumn{1}{c|}{21}                     & \multicolumn{1}{c|}{24}                      & 45             \\ \hline
\multicolumn{1}{|r|}{\textbf{Never taken}}         & \multicolumn{1}{c|}{2}                      & \multicolumn{1}{c|}{1}                       & 3              \\ \hline
\multicolumn{1}{|r|}{\textbf{Declined to answer}}  & \multicolumn{1}{c|}{3}                      & \multicolumn{1}{c|}{2}                       & 5              \\ \hline
\end{tabular}

%% file: tables/cwes.tex
\begin{tabular}{|l|l|ccc|ccc|}
\hline
\multicolumn{1}{|c|}{\multirow{2}{*}{\textbf{Function}}}                                                  & \multicolumn{1}{c|}{\multirow{2}{*}{\textbf{Group}}} & \multicolumn{3}{c|}{\textbf{Compiling}}                                                 & \multicolumn{3}{c|}{\textbf{Passing}}                                                   \\ \cline{3-8} 
\multicolumn{1}{|c|}{}                                                                                    & \multicolumn{1}{c|}{}                                & \multicolumn{1}{c|}{\textbf{N}} & \multicolumn{1}{c|}{\textbf{\# CWEs}} & \textbf{Rate} & \multicolumn{1}{c|}{\textbf{N}} & \multicolumn{1}{c|}{\textbf{\# CWEs}} & \textbf{Rate} \\ \hline
\multirow{3}{*}{\begin{tabular}[c]{@{}l@{}}list\_add\_\\ item\_at\_pos\end{tabular}}                      & Control                                              & \multicolumn{1}{c|}{20}         & \multicolumn{1}{c|}{61}               & \cellcolor[HTML]{6195C9}3.05 $\dagger$         & \multicolumn{1}{c|}{12}         & \multicolumn{1}{c|}{38}               & 3.17          \\ \cline{2-8} 
                                                                                                          & Assisted                                             & \multicolumn{1}{c|}{26}         & \multicolumn{1}{c|}{88}               & \cellcolor[HTML]{6195C9}3.38 $\dagger$          & \multicolumn{1}{c|}{16}         & \multicolumn{1}{c|}{53}               & 3.31          \\ \cline{2-8} 
                                                                                                          & Autopilot                                            & \multicolumn{1}{c|}{5}          & \multicolumn{1}{c|}{13}               & 2.6           & \multicolumn{1}{c|}{1}          & \multicolumn{1}{c|}{2}                & 2.0           \\ \hline
\multirow{3}{*}{list\_cost\_sum}                                                                          & Control                                              & \multicolumn{1}{c|}{13}         & \multicolumn{1}{c|}{12}               & 0.92          & \multicolumn{1}{c|}{10}         & \multicolumn{1}{c|}{10}               & 1.0           \\ \cline{2-8} 
                                                                                                          & Assisted                                             & \multicolumn{1}{c|}{16}         & \multicolumn{1}{c|}{14}               & 0.88          & \multicolumn{1}{c|}{14}         & \multicolumn{1}{c|}{14}               & 1.0           \\ \cline{2-8} 
                                                                                                          & Autopilot                                            & \multicolumn{1}{c|}{5}          & \multicolumn{1}{c|}{9}                & 1.8           & \multicolumn{1}{c|}{4}          & \multicolumn{1}{c|}{8}                & 2.0           \\ \hline
\multirow{3}{*}{list\_deduplicate}                                                                        & Control                                              & \multicolumn{1}{c|}{12}         & \multicolumn{1}{c|}{20}               & \cellcolor[HTML]{FFCE93}1.67 $\dagger$          & \multicolumn{1}{c|}{4}          & \multicolumn{1}{c|}{7}                & 1.75          \\ \cline{2-8} 
                                                                                                          & Assisted                                             & \multicolumn{1}{c|}{15}         & \multicolumn{1}{c|}{14}               & \cellcolor[HTML]{FFCE93}0.93 $\dagger$         & \multicolumn{1}{c|}{3}          & \multicolumn{1}{c|}{5}                & 1.67          \\ \cline{2-8} 
                                                                                                          & Autopilot                                            & \multicolumn{1}{c|}{5}          & \multicolumn{1}{c|}{4}                & 0.8           & \multicolumn{1}{c|}{1}          & \multicolumn{1}{c|}{2}                & 2.0           \\ \hline
\multirow{3}{*}{\begin{tabular}[c]{@{}l@{}}list\_find\_\\ highest\_price\_\\ item\_position\end{tabular}} & Control                                              & \multicolumn{1}{c|}{14}         & \multicolumn{1}{c|}{14}               & \cellcolor[HTML]{FFCE93}1.00  $\dagger$          & \multicolumn{1}{c|}{8}          & \multicolumn{1}{c|}{11}               & \cellcolor[HTML]{FFCE93}1.38 $\dagger$        \\ \cline{2-8} 
                                                                                                          & Assisted                                             & \multicolumn{1}{c|}{19}         & \multicolumn{1}{c|}{12}               & \cellcolor[HTML]{FFCE93}0.63 $\dagger$          & \multicolumn{1}{c|}{11}         & \multicolumn{1}{c|}{09}               & \cellcolor[HTML]{FFCE93}0.82 $\dagger$         \\ \cline{2-8} 
                                                                                                          & Autopilot                                            & \multicolumn{1}{c|}{5}          & \multicolumn{1}{c|}{14}               & 2.8           & \multicolumn{1}{c|}{1}          & \multicolumn{1}{c|}{1}                & 1.0           \\ \hline
\multirow{3}{*}{\begin{tabular}[c]{@{}l@{}}list\_item\_\\ to\_string\end{tabular}}                        & Control                                              & \multicolumn{1}{c|}{21}         & \multicolumn{1}{c|}{47}               & 2.24          & \multicolumn{1}{c|}{13}         & \multicolumn{1}{c|}{29}               & 2.23          \\ \cline{2-8} 
                                                                                                          & Assisted                                             & \multicolumn{1}{c|}{26}         & \multicolumn{1}{c|}{56}               & 2.15          & \multicolumn{1}{c|}{20}         & \multicolumn{1}{c|}{43}               & 2.15          \\ \cline{2-8} 
                                                                                                          & Autopilot                                            & \multicolumn{1}{c|}{5}          & \multicolumn{1}{c|}{10}               & 2.0           & \multicolumn{1}{c|}{4}          & \multicolumn{1}{c|}{6}                & 1.5          \\ \hline
\multirow{3}{*}{list\_load}                                                                               & Control                                              & \multicolumn{1}{c|}{12}         & \multicolumn{1}{c|}{19}               & 1.58          & \multicolumn{1}{c|}{4}          & \multicolumn{1}{c|}{6}                & \cellcolor[HTML]{6195C9}1.5 $\dagger$           \\ \cline{2-8} 
                                                                                                          & Assisted                                             & \multicolumn{1}{c|}{17}         & \multicolumn{1}{c|}{27}               & 1.59           & \multicolumn{1}{c|}{4}          & \multicolumn{1}{c|}{8}                & \cellcolor[HTML]{6195C9}2.0 $\dagger$           \\ \cline{2-8} 
                                                                                                          & Autopilot                                            & \multicolumn{1}{c|}{5}          & \multicolumn{1}{c|}{0}                & 0.0           & \multicolumn{1}{c|}{0}          & \multicolumn{1}{c|}{0}                & \textit{}     \\ \hline
\multirow{3}{*}{list\_print}                                                                              & Control                                              & \multicolumn{1}{c|}{24}         & \multicolumn{1}{c|}{16}               & \cellcolor[HTML]{6195C9}0.67          & \multicolumn{1}{c|}{9}          & \multicolumn{1}{c|}{4}                & \cellcolor[HTML]{6195C9}0.44 $\dagger$           \\ \cline{2-8} 
                                                                                                          & Assisted                                             & \multicolumn{1}{c|}{27}         & \multicolumn{1}{c|}{22}               & \cellcolor[HTML]{6195C9}0.81           & \multicolumn{1}{c|}{13}         & \multicolumn{1}{c|}{10}               & \cellcolor[HTML]{6195C9}0.77 $\dagger$          \\ \cline{2-8} 
                                                                                                          & Autopilot                                            & \multicolumn{1}{c|}{5}          & \multicolumn{1}{c|}{2}                & 0.4           & \multicolumn{1}{c|}{1}          & \multicolumn{1}{c|}{1}                & 1.0           \\ \hline
\multirow{3}{*}{\begin{tabular}[c]{@{}l@{}}list\_remove\_\\ item\_at\_pos\end{tabular}}                   & Control                                              & \multicolumn{1}{c|}{13}         & \multicolumn{1}{c|}{49}               & 3.77          & \multicolumn{1}{c|}{10}          & \multicolumn{1}{c|}{43}               & 4.30 $\dagger$          \\ \cline{2-8} 
                                                                                                          & Assisted                                             & \multicolumn{1}{c|}{19}         & \multicolumn{1}{c|}{74}               & 3.89          & \multicolumn{1}{c|}{14}         & \multicolumn{1}{c|}{51}               & 3.92 $\dagger$         \\ \cline{2-8} 
                                                                                                          & Autopilot                                            & \multicolumn{1}{c|}{5}          & \multicolumn{1}{c|}{16}               & 3.2           & \multicolumn{1}{c|}{3}          & \multicolumn{1}{c|}{9}                & 3.00            \\ \hline
\multirow{3}{*}{list\_save}                                                                               & Control                                              & \multicolumn{1}{c|}{14}         & \multicolumn{1}{c|}{4}                & 0.29          & \multicolumn{1}{c|}{7}          & \multicolumn{1}{c|}{1}                & \cellcolor[HTML]{6195C9}0.14          \\ \cline{2-8} 
                                                                                                          & Assisted                                             & \multicolumn{1}{c|}{17}         & \multicolumn{1}{c|}{5}                & 0.29          & \multicolumn{1}{c|}{7}          & \multicolumn{1}{c|}{2}                & \cellcolor[HTML]{6195C9}0.29          \\ \cline{2-8} 
                                                                                                          & Autopilot                                            & \multicolumn{1}{c|}{5}          & \multicolumn{1}{c|}{0}                & 0.0           & \multicolumn{1}{c|}{0}          & \multicolumn{1}{c|}{0}                & \textit{}     \\ \hline
\multirow{3}{*}{\begin{tabular}[c]{@{}l@{}}list\_swap\_\\ item\_positions\end{tabular}}                   & Control                                              & \multicolumn{1}{c|}{12}         & \multicolumn{1}{c|}{26}               & 2.17           & \multicolumn{1}{c|}{5}          & \multicolumn{1}{c|}{11}                & \cellcolor[HTML]{6195C9}2.20 $\dagger$           \\ \cline{2-8} 
                                                                                                          & Assisted                                             & \multicolumn{1}{c|}{20}         & \multicolumn{1}{c|}{43}               & 2.15          & \multicolumn{1}{c|}{6}          & \multicolumn{1}{c|}{7}               & \cellcolor[HTML]{6195C9}1.17 $\dagger$         \\ \cline{2-8} 
                                                                                                          & Autopilot                                            & \multicolumn{1}{c|}{5}          & \multicolumn{1}{c|}{15}               & 3.0           & \multicolumn{1}{c|}{0}          & \multicolumn{1}{c|}{0}                & \textit{}     \\ \hline
\multirow{3}{*}{\begin{tabular}[c]{@{}l@{}}list\_update\_\\ item\_at\_pos\end{tabular}}                   & Control                                              & \multicolumn{1}{c|}{14}         & \multicolumn{1}{c|}{38}               & \cellcolor[HTML]{6195C9}2.71 $\dagger$         & \multicolumn{1}{c|}{10}          & \multicolumn{1}{c|}{36}               & 3.6          \\ \cline{2-8} 
                                                                                                          & Assisted                                             & \multicolumn{1}{c|}{24}         & \multicolumn{1}{c|}{88}               & \cellcolor[HTML]{6195C9}3.67 $\dagger$         & \multicolumn{1}{c|}{16}         & \multicolumn{1}{c|}{60}               & 3.75          \\ \cline{2-8} 
                                                                                                          & Autopilot                                            & \multicolumn{1}{c|}{5}          & \multicolumn{1}{c|}{16}               & 3.2           & \multicolumn{1}{c|}{3}          & \multicolumn{1}{c|}{13}               & 4.33          \\ \hline
\multirow{3}{*}{\textbf{Totals}}                                                                          & \textbf{Control}                                     & \multicolumn{1}{c|}{139}        & \multicolumn{1}{c|}{290}              & 2.09          & \multicolumn{1}{c|}{84}         & \multicolumn{1}{c|}{180}              & 2.14          \\ \cline{2-8} 
                                                                                                          & \textbf{Assisted}                                    & \multicolumn{1}{c|}{204}        & \multicolumn{1}{c|}{451}              & 2.21          & \multicolumn{1}{c|}{124}        & \multicolumn{1}{c|}{278}              & 2.24          \\ \cline{2-8} 
                                                                                                          & \textbf{Autopilot}                                   & \multicolumn{1}{c|}{49}         & \multicolumn{1}{c|}{99}               & 2.02          & \multicolumn{1}{c|}{18}         & \multicolumn{1}{c|}{43}               & 2.39          \\ \hline
\end{tabular}

%% file: tables/bug-origins.tex
\begin{tabular}{|c|c|c|c|c|}
\hline
\textbf{\begin{tabular}[c]{@{}c@{}}Participant\\ UUID\end{tabular}} & \textbf{\begin{tabular}[c]{@{}c@{}}First location\\ of bug\\ (document /\\ suggestion)\end{tabular}} & \textbf{\begin{tabular}[c]{@{}c@{}}\# Bug\\ suggestions\end{tabular}} & \textbf{\begin{tabular}[c]{@{}c@{}}\# Bug\\ suggestions\\ accepted\end{tabular}} & \textbf{\begin{tabular}[c]{@{}c@{}}\# Bugs\\ in final\\ file\end{tabular}} \\ \hline
0640                                                              & Suggestion                                                                                           & 5                                                                     & 3                                                                                & 3                                                                          \\ \hline
1f1c                                                              & Document                                                                                             & 5                                                                     & 0                                                                                & 2                                                                          \\ \hline
2125                                                              & Document                                                                                             & 0                                                                     & 0                                                                                & 3                                                                          \\ \hline
26a4                                                              & Suggestion                                                                                           & 3                                                                     & 1                                                                                & 2                                                                          \\ \hline
3533                                                              & Suggestion                                                                                           & 2                                                                     & 1                                                                                & 1                                                                          \\ \hline
36de                                                              & Suggestion                                                                                           & 69                                                                    & 5                                                                                & 4                                                                          \\ \hline
3cff                                                              & Suggestion                                                                                           & 2                                                                     & 2                                                                                & 2                                                                          \\ \hline
514e                                                              & Document                                                                                             & 1                                                                     & 1                                                                                & 1                                                                          \\ \hline
7193                                                              & Suggestion                                                                                           & 13                                                                    & 1                                                                                & 2                                                                          \\ \hline
74bd                                                              & Suggestion                                                                                           & 4                                                                     & 2                                                                                & 2                                                                          \\ \hline
925c                                                              & Suggestion                                                                                           & 8                                                                     & 2                                                                                & 1                                                                          \\ \hline
a3ed                                                              & Suggestion                                                                                           & 10                                                                    & 2                                                                                & 2                                                                          \\ \hline
a4b3                                                              & Suggestion                                                                                           & 11                                                                    & 5                                                                                & 4                                                                          \\ \hline
a5ba                                                              & Document                                                                                             & 0                                                                     & 0                                                                                & 1                                                                          \\ \hline
a80d                                                              & Document                                                                                             & 6                                                                     & 3                                                                                & 3                                                                          \\ \hline
a974                                                              & Suggestion                                                                                           & 12                                                                    & 5                                                                                & 3                                                                          \\ \hline
b59f                                                              & Suggestion                                                                                           & 8                                                                     & 2                                                                                & 2                                                                          \\ \hline
be6f                                                              & Suggestion                                                                                           & 4                                                                     & 1                                                                                & 2                                                                          \\ \hline
c23b                                                              & Suggestion                                                                                           & 20                                                                    & 10                                                                               & 5                                                                          \\ \hline
dac3                                                              & Document                                                                                             & 10                                                                    & 2                                                                                & 2                                                                          \\ \hline
dc47                                                              & Suggestion                                                                                           & 1                                                                     & 0                                                                                & 2                                                                          \\ \hline
ddac                                                              & Suggestion                                                                                           & 13                                                                    & 1                                                                                & 1                                                                          \\ \hline
ec83                                                              & Document                                                                                             & 11                                                                    & 3                                                                                & 2                                                                          \\ \hline
fd62                                                              & Suggestion                                                                                           & 12                                                                    & 1                                                                                & 1                                                                           \\ \hline
\end{tabular}

%% file: sections/05Discussion.tex
\section{Discussion}

\subsection{Implications for LLM assistants}

\textbf{Functionality (RQ1):} our results corroborate recent studies that have suggested that LLM assistants
improve developer productivity~\cite{ziegler_productivity_2022,tabachnyk_ml-enhanced_2022}. \textbf{While we do not directly measure productivity, the fact that `assisted' users submitted more lines of code and completed a greater fraction of functions suggests enhanced productivity.}
One surprising result was the relatively high quality of code produced in `autopilot' mode (albeit for a relatively simple task).

\textbf{Security (RQ2):} \newt{While prior work found that LLM code assistants may suggest security-critical bugs/CWEs~\cite{pearce_asleep_2022}, it did not attempt to determine a comparison of the tool against human developers nor did it examine how the (potentially vulnerable) suggestions may impact developers using the tools.
Meanwhile, other studies which have included human developers~\cite{imai_is_2022,ziegler_productivity_2022,tabachnyk_ml-enhanced_2022} have not considered security.
As such, to the authors knowledge, the user study presented in this work is the first such study which measures how LLM suggestions may impact the security of the code.} We have found no conclusive evidence to support the claim LLM assistants \emph{increase} CWE incidence in code in general, even when we looked only at severe CWEs. Our results indicate that the security impact in this setting is small: AI-assisted users produce critical security bugs at a rate no greater than 10\% higher than the control, indicating that LLMs do not introduce new security risks. 
\textbf{This suggests that security concerns with LLM assistants might not be as severe as initially suggested, although studies with larger sample sizes and diverse user groups are warranted.}

\textbf{Bug origins (RQ3):}
Our results indicate that users interact with the LLM in interesting ways.
Users provide prompts which may include bugs, accept buggy prompts which end up in the `completed' programs as well as accept bugs which are later removed. 
In some cases, users also end up with more bugs than were suggested by the LLM!
In addition, the users that accepted the most bugs from the LLM also had the most bugs in their final files, further suggesting that the use of a buggy LLM may lead users toward buggy code.

\subsection{Threats to validity}

\textbf{User selection:} This study recruited university students rather than professional developers. \newt{While this may have an impact on the generalizability of the results due to differences in behavior and code performance,} previous work~\cite{acar_security_2017} has found no difference between experienced software developers and students regarding security-aware coding. In this work, we observed a defect density of 0.15 bugs/LoC which while greater than reported figures of 0.07 bugs/LoC~\cite{assaraf_this_2015}, is reasonable given the time constraints of the assignment.

\textbf{Code assignment difficulty:} We designed both the assignment and chose the programming language with the intention of examining how the developers might miss bugs in their designs. 
As such, the singly-linked shopping list has a number of unusual traits and a non-optimal API. 
This increases the difficulty of the assignment, which may itself have an impact on the study results---if a developer is unable to `solve' the coding challenge at hand, they may get frustrated and hand in a substandard solution.
Further, C is considered a more difficult programming language for inexperienced programmers than other languages~\cite{fangohr_comparison_2004} such as Python or MATLAB. 
\newt{Other contexts (i.e. other languages, other programming tasks) may yield different results than those found in this study.}

\textbf{Data capture:} Due to limitations of the cloud-based IDE, it was not possible to capture all data from participants.
For instance, rather than capturing every keypress from the users, we were restricted to taking `snapshots' of their development over time (every 60 seconds). 
This limits the kind of fine-grained analysis that might have been possible with more pervasive measurements.

%% file: sections/09Conclusion.tex
\section{Conclusions}

In this paper we set out to investigate the cybersecurity impact of LLM code suggestions on participants writing code in a user study.
With N=58 users, we determined that the LLM has a likely beneficial impact on functional correctness; and does not increase the incidence rates of severe security bugs in our context \newt{(i.e., low level C code with pointer and array manipulations)}. %
This is somewhat surprising given the existing published studies on how vulnerable code can be suggested by the LLMs~\cite{pearce_asleep_2022,siddiq_empirical_2022}.
When considering the origin of bugs that were found, the data suggests that the users do not use the extra productivity benefits to fix bugs in their code---although suggestions are being modified (e.g. variable names), if a suggestion contained a bug it may not be fixed. This suggests that further research needs to be undertaken on highlighting problematic lines of code (`nutritional labels') to encourage users to check for security in real-time, as well as improving code LLMs so that they can produce code that is \emph{more} secure than the user's existing code. %

%% file: sections/10Appendix.tex
\section*{Appendix}

\newt{\textbf{User study recruitment and ethical considerations:}

This study involved human participants and was approved by New York University's Institution Review Board (IRB) as \#IRB-FY2022-6074. The key details are noted here.

Participants were recruited in phases from students and ex-students of the classes of two of the authors of this paper.
As there is a potential power dynamic between an instructor and their students, a hard firewall was established between student participants and their instructors. All knowledge of enrolled participants was restricted to a single research investigator (and author of this paper), and this investigator was not the instructor or supervisor of any of the study participants. Students were informed of this firewall, and were informed that their participation status would be kept strictly confidential from their instructors, and that participating (or not participating) would have no impact on their grades. 

During recruitment, participants were told both verbally and via advertising material for the study that they would be randomly divided into two groups, one with access to an LLM code-assistant, and one without, and would then be asked to complete a programming challenge. They were also told that the code between each of the groups would be compared (Quote from the advertising material: ``Will one group outperform the other?'', see \autoref{fig:recruitment-text}). Verbal discussion of the study included potential metrics, including time taken, functionality measurements, and security issues. 

As participants were informed that metrics including security analysis would be a part of the study, this study did not deceive any participants. 
It also did not deceive them with regards to the LLM itself, as AI responses from the Codex LLM were not modified (e.g. no bugs were artificially added to suggestions above the ones already present).}

\begin{figure}[H]
    \centering
    \noindent\fbox{%
    \parbox{0.95\linewidth}{%
    Large Language Models such as GitHub Copilot and OpenAI Codex have recently been commercially released with the goal of \textit{helping developers write software code}. While the marketing material touts the benefits of these ``AI Pair Programmers'', the actual impacts of these LLMs is yet to be formally investigated. In this research, we aim to begin this exploration by challenging participants such as yourself with completing a range of programming questions similar to those posed at the undergraduate level in computer science and software engineering courses. The research question is simple: Half the participants will have assistance via an LLM (OpenAI Codex) and the other half will not. \textbf{Will one group outperform the other?}
    }%
}
    \caption{Introductory paragraph to the recruitment material.}
    \label{fig:recruitment-text}
    \vspace{-4mm}
\end{figure}

\begin{figure}[h]
\includegraphics[width=\columnwidth]{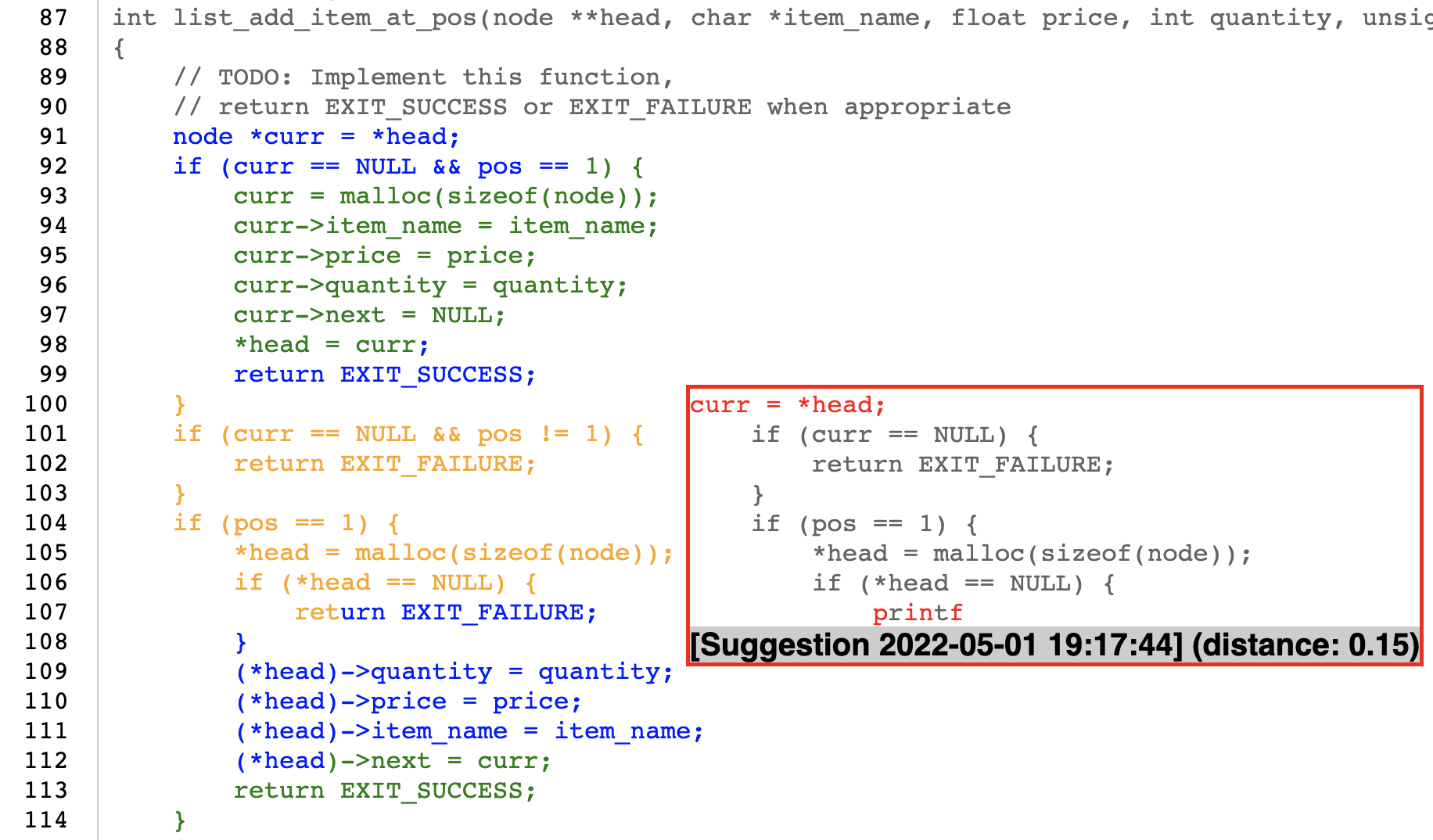}
\caption{Our visualization tool explores LLM suggestion acceptance. Grey: Initial template; blue: human-written code; green: code accepted exactly from Codex suggestion; orange: approximate matches. Pop-up: Codex suggestion (on hover).}
\label{fig:suggestion_viz}
\end{figure}

\textbf{CWE frequency within each study group:}

\autoref{table:cwe-list} lists the most common CWEs from each study group with their descriptions, downstream-CWEs if they are severe, and the MITRE `Top 25' rank.

\autoref{tbl:severe-cwe-counts-per-function} presents the severe CWE counts per function by study group, with bugs associated according to \autoref{table:cwe-list}.
The N for each category refers to the (N)umber of compiling functions from participants. Rate refers to the count of this CWE divided by the N. The `Autopilot' group contains only to the first 5 answers from \texttt{code-cushman-001}.

\begin{table}[H]
    \caption{Top 10 most common CWEs in each study group, along with downstream severe CWEs if a non-severe CWE would lead to a different severe CWE.}
    \label{table:cwe-list}
    \resizebox{\linewidth}{!}{\input{tables/cwes-descriptions}}
    \end{table}

\begin{figure}[h]
\includegraphics[width=\columnwidth]{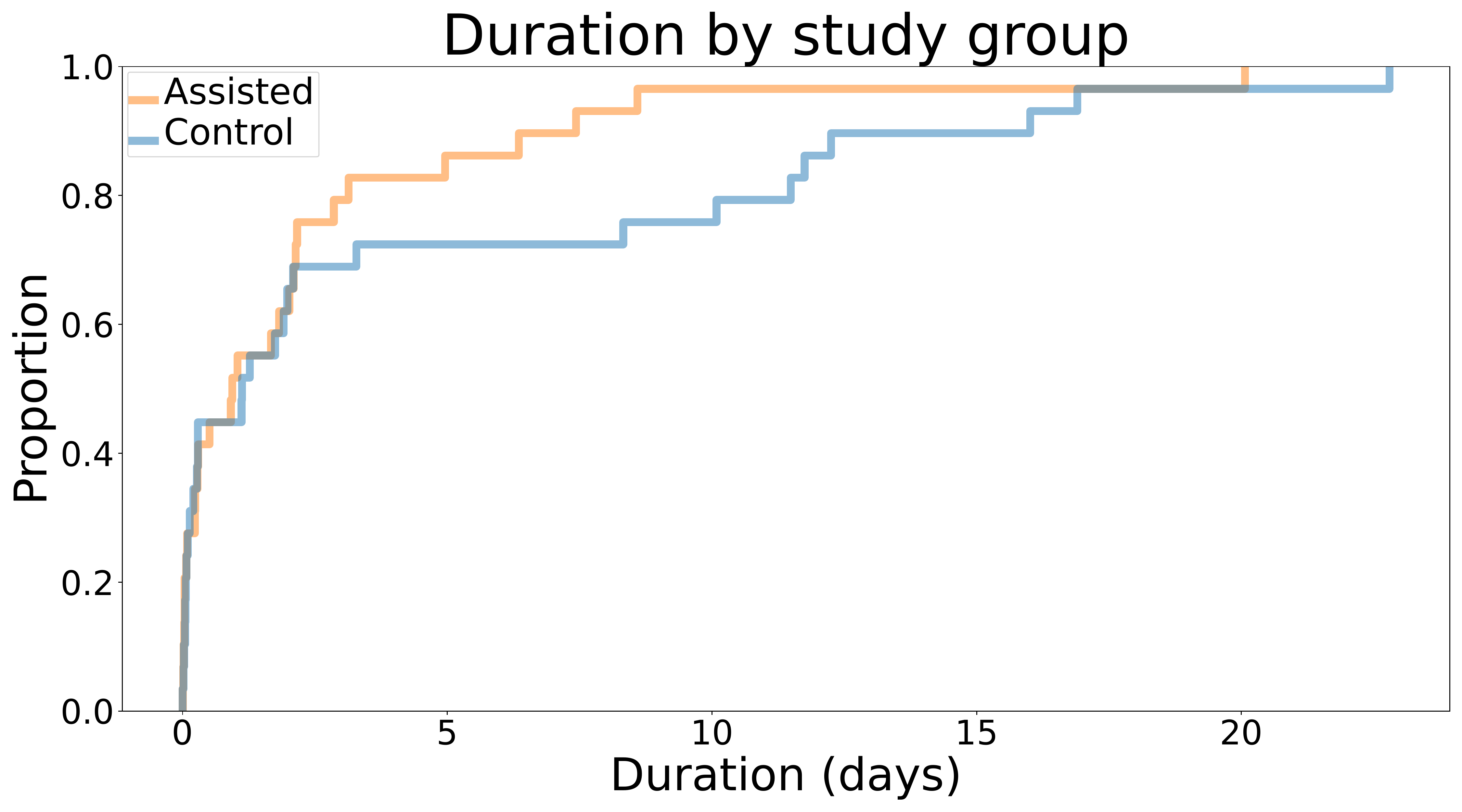}
\caption{\newt{Study completion time between `Assisted' and `Control' participant groups. Although the deadline was 14 days, a small minority of participants from both groups required additional time.}}
\label{fig:study-duration}
\end{figure}

\begin{table*}[t]

\centering
\vspace{-1mm}
\caption{Severe CWE counts per function by study group. `Autopilot' group refers to the first 5 \texttt{code-cushman-001} answers.\label{tbl:severe-cwe-counts-per-function}}
  \centering
\resizebox{0.83\textwidth}{!}{\input{tables/cwes-per-f}}
\end{table*}

%% file: tables/cwes-descriptions.tex
\begin{tabular}{|l|l|l|}
\hline
\textbf{CWE ID} & \textbf{Description}                      & \textbf{\begin{tabular}[c]{@{}l@{}}`Top 25' \\ Rank\end{tabular}} \\ \hline
CWE-476         & NULL Pointer Dereference                  & 11                                                                \\ \hline
CWE-758         & Reliance on Undefined Behavior            & -                                                                 \\ \hline
CWE-401         & Missing Release of Memory                 &                                                                   \\
$\hookrightarrow$ CWE-400      & Uncontrolled Resource Consumption         & 23                                                                \\ \hline
CWE-252         & Unchecked Return Value            & -                                                                 \\ \hline
CWE-416         & Use after Free                            & 7                                                                 \\ \hline
CWE-787         & Out-of-bounds Write                       & 1                                                                 \\ \hline
CWE-843         & Access using Incompatible Type            &                                                                  \\
$\hookrightarrow$ CWE-119      & Improper Restriction of Buffer Ops & 19                                                                \\ \hline
CWE-457         & Use of Uninitialized Variable             &                                                                  \\
$\hookrightarrow$ CWE-119      & Improper Restriction of Buffer Operations & 19                                                                \\ \hline
CWE-835         & Infinite Loop        & -                                                                 \\ \hline
\end{tabular}

%% file: tables/cwes-per-f.tex
\begin{tabular}{|l|l|l|lll|lll|}
\hline
\multicolumn{1}{|c|}{\multirow{2}{*}{\textbf{Function Name}}} & \multicolumn{1}{c|}{\multirow{2}{*}{\textbf{Group}}} & \multicolumn{1}{c|}{\multirow{2}{*}{\textbf{Observed CWE}}}                      & \multicolumn{3}{c|}{\textbf{Compiling}}                                                                                                                    & \multicolumn{3}{c|}{\textbf{Passing}}                                                                                                                      \\ \cline{4-9} 
\multicolumn{1}{|c|}{}                                        & \multicolumn{1}{c|}{}                                & \multicolumn{1}{c|}{}                                                            & \multicolumn{1}{c|}{\textbf{N}} & \multicolumn{1}{c|}{\textbf{\begin{tabular}[c]{@{}c@{}}\# this\\ CWE\end{tabular}}} & \multicolumn{1}{c|}{\textbf{Rate}} & \multicolumn{1}{c|}{\textbf{N}} & \multicolumn{1}{c|}{\textbf{\begin{tabular}[c]{@{}c@{}}\# this\\ CWE\end{tabular}}} & \multicolumn{1}{c|}{\textbf{Rate}} \\ \hline
\multirow{11}{*}{list\_add\_item\_at\_pos}                    & \multirow{4}{*}{Control}                             & CWE-119: Improper Restriction of Operations within the Bounds of a Memory Buffer & \multicolumn{1}{l|}{20}         & \multicolumn{1}{l|}{7}                                                              & 0.35                               & \multicolumn{1}{l|}{12}         & \multicolumn{1}{l|}{2}                                                              & 0.17                               \\ \cline{3-9} 
                                                              &                                                      & CWE-400: Uncontrolled Resource Consumption                                       & \multicolumn{1}{l|}{20}         & \multicolumn{1}{l|}{9}                                                              & 0.45                               & \multicolumn{1}{l|}{12}         & \multicolumn{1}{l|}{6}                                                              & 0.5                                \\ \cline{3-9} 
                                                              &                                                      & CWE-416: Use After Free                                                          & \multicolumn{1}{l|}{20}         & \multicolumn{1}{l|}{19}                                                             & 0.95                               & \multicolumn{1}{l|}{12}         & \multicolumn{1}{l|}{14}                                                             & 1.17                               \\ \cline{3-9} 
                                                              &                                                      & CWE-476: NULL Pointer Dereference                                                & \multicolumn{1}{l|}{20}         & \multicolumn{1}{l|}{26}                                                             & 1.3                                & \multicolumn{1}{l|}{12}         & \multicolumn{1}{l|}{16}                                                             & 1.33                               \\ \cline{2-9} 
                                                              & \multirow{4}{*}{Assisted}                            & CWE-119: Improper Restriction of Operations within the Bounds of a Memory Buffer & \multicolumn{1}{l|}{26}         & \multicolumn{1}{l|}{13}                                                             & 0.5                                & \multicolumn{1}{l|}{16}         & \multicolumn{1}{l|}{7}                                                              & 0.44                               \\ \cline{3-9} 
                                                              &                                                      & CWE-400: Uncontrolled Resource Consumption                                       & \multicolumn{1}{l|}{26}         & \multicolumn{1}{l|}{14}                                                             & 0.54                               & \multicolumn{1}{l|}{16}         & \multicolumn{1}{l|}{7}                                                              & 0.44                               \\ \cline{3-9} 
                                                              &                                                      & CWE-416: Use After Free                                                          & \multicolumn{1}{l|}{26}         & \multicolumn{1}{l|}{30}                                                             & 1.15                               & \multicolumn{1}{l|}{16}         & \multicolumn{1}{l|}{22}                                                             & 1.38                               \\ \cline{3-9} 
                                                              &                                                      & CWE-476: NULL Pointer Dereference                                                & \multicolumn{1}{l|}{26}         & \multicolumn{1}{l|}{31}                                                             & 1.19                               & \multicolumn{1}{l|}{16}         & \multicolumn{1}{l|}{17}                                                             & 1.06                               \\ \cline{2-9} 
                                                              & \multirow{3}{*}{Autopilot}                           & CWE-400: Uncontrolled Resource Consumption                                       & \multicolumn{1}{l|}{5}          & \multicolumn{1}{l|}{2}                                                              & 0.4                                & \multicolumn{1}{l|}{1}          & \multicolumn{1}{l|}{0}                                                              & 0.0                                \\ \cline{3-9} 
                                                              &                                                      & CWE-416: Use After Free                                                          & \multicolumn{1}{l|}{5}          & \multicolumn{1}{l|}{4}                                                              & 0.8                                & \multicolumn{1}{l|}{1}          & \multicolumn{1}{l|}{0}                                                              & 0.0                                \\ \cline{3-9} 
                                                              &                                                      & CWE-476: NULL Pointer Dereference                                                & \multicolumn{1}{l|}{5}          & \multicolumn{1}{l|}{7}                                                              & 1.4                                & \multicolumn{1}{l|}{1}          & \multicolumn{1}{l|}{2}                                                              & 2.0                                \\ \hline
\multirow{3}{*}{list\_cost\_sum}                              & Control                                              & CWE-476: NULL Pointer Dereference                                                & \multicolumn{1}{l|}{13}         & \multicolumn{1}{l|}{12}                                                             & 0.92                               & \multicolumn{1}{l|}{10}         & \multicolumn{1}{l|}{10}                                                             & 1.0                                \\ \cline{2-9} 
                                                              & Assisted                                             & CWE-476: NULL Pointer Dereference                                                & \multicolumn{1}{l|}{16}         & \multicolumn{1}{l|}{14}                                                             & 0.88                               & \multicolumn{1}{l|}{14}         & \multicolumn{1}{l|}{14}                                                             & 1.0                                \\ \cline{2-9} 
                                                              & Autopilot                                            & CWE-476: NULL Pointer Dereference                                                & \multicolumn{1}{l|}{5}          & \multicolumn{1}{l|}{9}                                                              & 1.8                                & \multicolumn{1}{l|}{4}          & \multicolumn{1}{l|}{8}                                                              & 2.0                                \\ \hline
\multirow{11}{*}{list\_deduplicate}                           & \multirow{5}{*}{Control}                             & CWE-119: Improper Restriction of Operations within the Bounds of a Memory Buffer & \multicolumn{1}{l|}{12}         & \multicolumn{1}{l|}{1}                                                              & 0.08                               & \multicolumn{1}{l|}{4}          & \multicolumn{1}{l|}{0}                                                              & 0.0                                \\ \cline{3-9} 
                                                              &                                                      & CWE-400: Uncontrolled Resource Consumption                                       & \multicolumn{1}{l|}{12}         & \multicolumn{1}{l|}{6}                                                              & 0.5                                & \multicolumn{1}{l|}{4}          & \multicolumn{1}{l|}{2}                                                              & 0.5                                \\ \cline{3-9} 
                                                              &                                                      & CWE-416: Use After Free                                                          & \multicolumn{1}{l|}{12}         & \multicolumn{1}{l|}{1}                                                              & 0.08                               & \multicolumn{1}{l|}{4}          & \multicolumn{1}{l|}{1}                                                              & 0.25                               \\ \cline{3-9} 
                                                              &                                                      & CWE-476: NULL Pointer Dereference                                                & \multicolumn{1}{l|}{12}         & \multicolumn{1}{l|}{11}                                                             & 0.92                               & \multicolumn{1}{l|}{4}          & \multicolumn{1}{l|}{4}                                                              & 1.0                                \\ \cline{3-9} 
                                                              &                                                      & CWE-787: Out-of-bounds Write                                                     & \multicolumn{1}{l|}{12}         & \multicolumn{1}{l|}{1}                                                              & 0.08                               & \multicolumn{1}{l|}{4}          & \multicolumn{1}{l|}{0}                                                              & 0.0                                \\ \cline{2-9} 
                                                              & \multirow{4}{*}{Assisted}                            & CWE-119: Improper Restriction of Operations within the Bounds of a Memory Buffer & \multicolumn{1}{l|}{16}         & \multicolumn{1}{l|}{1}                                                              & 0.06                               & \multicolumn{1}{l|}{3}          & \multicolumn{1}{l|}{0}                                                              & 0.0                                \\ \cline{3-9} 
                                                              &                                                      & CWE-400: Uncontrolled Resource Consumption                                       & \multicolumn{1}{l|}{16}         & \multicolumn{1}{l|}{4}                                                              & 0.25                               & \multicolumn{1}{l|}{3}          & \multicolumn{1}{l|}{4}                                                              & 1.33                               \\ \cline{3-9} 
                                                              &                                                      & CWE-416: Use After Free                                                          & \multicolumn{1}{l|}{16}         & \multicolumn{1}{l|}{2}                                                              & 0.13                               & \multicolumn{1}{l|}{3}          & \multicolumn{1}{l|}{0}                                                              & 0.0                                \\ \cline{3-9} 
                                                              &                                                      & CWE-476: NULL Pointer Dereference                                                & \multicolumn{1}{l|}{16}         & \multicolumn{1}{l|}{7}                                                              & 0.44                               & \multicolumn{1}{l|}{3}          & \multicolumn{1}{l|}{1}                                                              & 0.33                               \\ \cline{2-9} 
                                                              & \multirow{2}{*}{Autopilot}                           & CWE-400: Uncontrolled Resource Consumption                                       & \multicolumn{1}{l|}{5}          & \multicolumn{1}{l|}{3}                                                              & 0.6                                & \multicolumn{1}{l|}{1}          & \multicolumn{1}{l|}{1}                                                              & 1.0                                \\ \cline{3-9} 
                                                              &                                                      & CWE-476: NULL Pointer Dereference                                                & \multicolumn{1}{l|}{5}          & \multicolumn{1}{l|}{1}                                                              & 0.2                                & \multicolumn{1}{l|}{1}          & \multicolumn{1}{l|}{1}                                                              & 1.0                                \\ \hline
\multirow{4}{*}{list\_find\_highest\_price\_item\_position}   & \multirow{2}{*}{Control}                             & CWE-119: Improper Restriction of Operations within the Bounds of a Memory Buffer & \multicolumn{1}{l|}{14}         & \multicolumn{1}{l|}{1}                                                              & 0.07                               & \multicolumn{1}{l|}{8}          & \multicolumn{1}{l|}{0}                                                              & 0.0                                \\ \cline{3-9} 
                                                              &                                                      & CWE-476: NULL Pointer Dereference                                                & \multicolumn{1}{l|}{14}         & \multicolumn{1}{l|}{13}                                                             & 0.93                               & \multicolumn{1}{l|}{8}          & \multicolumn{1}{l|}{11}                                                             & 1.38                               \\ \cline{2-9} 
                                                              & Assisted                                             & CWE-476: NULL Pointer Dereference                                                & \multicolumn{1}{l|}{19}         & \multicolumn{1}{l|}{12}                                                             & 0.63                               & \multicolumn{1}{l|}{11}         & \multicolumn{1}{l|}{9}                                                              & 0.82                               \\ \cline{2-9} 
                                                              & Autopilot                                            & CWE-476: NULL Pointer Dereference                                                & \multicolumn{1}{l|}{5}          & \multicolumn{1}{l|}{14}                                                             & 2.8                                & \multicolumn{1}{l|}{1}          & \multicolumn{1}{l|}{1}                                                              & 1.0                                \\ \hline
\multirow{9}{*}{list\_item\_to\_string}                       & \multirow{3}{*}{Control}                             & CWE-119: Improper Restriction of Operations within the Bounds of a Memory Buffer & \multicolumn{1}{l|}{21}         & \multicolumn{1}{l|}{4}                                                              & 0.19                               & \multicolumn{1}{l|}{13}         & \multicolumn{1}{l|}{0}                                                              & 0.0                                \\ \cline{3-9} 
                                                              &                                                      & CWE-476: NULL Pointer Dereference                                                & \multicolumn{1}{l|}{21}         & \multicolumn{1}{l|}{27}                                                             & 1.29                               & \multicolumn{1}{l|}{13}         & \multicolumn{1}{l|}{17}                                                             & 1.31                               \\ \cline{3-9} 
                                                              &                                                      & CWE-787: Out-of-bounds Write                                                     & \multicolumn{1}{l|}{21}         & \multicolumn{1}{l|}{16}                                                             & 0.76                               & \multicolumn{1}{l|}{13}         & \multicolumn{1}{l|}{12}                                                             & 0.92                               \\ \cline{2-9} 
                                                              & \multirow{4}{*}{Assisted}                            & CWE-119: Improper Restriction of Operations within the Bounds of a Memory Buffer & \multicolumn{1}{l|}{26}         & \multicolumn{1}{l|}{2}                                                              & 0.08                               & \multicolumn{1}{l|}{20}         & \multicolumn{1}{l|}{1}                                                              & 0.05                               \\ \cline{3-9} 
                                                              &                                                      & CWE-416: Use After Free                                                          & \multicolumn{1}{l|}{26}         & \multicolumn{1}{l|}{1}                                                              & 0.04                               & \multicolumn{1}{l|}{20}         & \multicolumn{1}{l|}{0}                                                              & 0.0                                \\ \cline{3-9} 
                                                              &                                                      & CWE-476: NULL Pointer Dereference                                                & \multicolumn{1}{l|}{26}         & \multicolumn{1}{l|}{27}                                                             & 1.04                               & \multicolumn{1}{l|}{20}         & \multicolumn{1}{l|}{22}                                                             & 1.1                                \\ \cline{3-9} 
                                                              &                                                      & CWE-787: Out-of-bounds Write                                                     & \multicolumn{1}{l|}{26}         & \multicolumn{1}{l|}{26}                                                             & 1.0                                & \multicolumn{1}{l|}{20}         & \multicolumn{1}{l|}{20}                                                             & 1.0                                \\ \cline{2-9} 
                                                              & \multirow{2}{*}{Autopilot}                           & CWE-476: NULL Pointer Dereference                                                & \multicolumn{1}{l|}{5}          & \multicolumn{1}{l|}{7}                                                              & 1.4                                & \multicolumn{1}{l|}{4}          & \multicolumn{1}{l|}{4}                                                              & 1.0                                \\ \cline{3-9} 
                                                              &                                                      & CWE-787: Out-of-bounds Write                                                     & \multicolumn{1}{l|}{5}          & \multicolumn{1}{l|}{3}                                                              & 0.6                                & \multicolumn{1}{l|}{4}          & \multicolumn{1}{l|}{3}                                                              & 0.75                               \\ \hline
\multirow{9}{*}{list\_load}                                   & \multirow{4}{*}{Control}                             & CWE-119: Improper Restriction of Operations within the Bounds of a Memory Buffer & \multicolumn{1}{l|}{12}         & \multicolumn{1}{l|}{3}                                                              & 0.25                               & \multicolumn{1}{l|}{4}          & \multicolumn{1}{l|}{0}                                                              & 0.0                                \\ \cline{3-9} 
                                                              &                                                      & CWE-400: Uncontrolled Resource Consumption                                       & \multicolumn{1}{l|}{12}         & \multicolumn{1}{l|}{5}                                                              & 0.42                               & \multicolumn{1}{l|}{4}          & \multicolumn{1}{l|}{1}                                                              & 0.25                               \\ \cline{3-9} 
                                                              &                                                      & CWE-476: NULL Pointer Dereference                                                & \multicolumn{1}{l|}{12}         & \multicolumn{1}{l|}{6}                                                              & 0.5                                & \multicolumn{1}{l|}{4}          & \multicolumn{1}{l|}{3}                                                              & 0.75                               \\ \cline{3-9} 
                                                              &                                                      & CWE-787: Out-of-bounds Write                                                     & \multicolumn{1}{l|}{12}         & \multicolumn{1}{l|}{5}                                                              & 0.42                               & \multicolumn{1}{l|}{4}          & \multicolumn{1}{l|}{2}                                                              & 0.5                                \\ \cline{2-9} 
                                                              & \multirow{5}{*}{Assisted}                            & CWE-119: Improper Restriction of Operations within the Bounds of a Memory Buffer & \multicolumn{1}{l|}{18}         & \multicolumn{1}{l|}{6}                                                              & 0.33                               & \multicolumn{1}{l|}{4}          & \multicolumn{1}{l|}{1}                                                              & 0.25                               \\ \cline{3-9} 
                                                              &                                                      & CWE-400: Uncontrolled Resource Consumption                                       & \multicolumn{1}{l|}{18}         & \multicolumn{1}{l|}{8}                                                              & 0.44                               & \multicolumn{1}{l|}{4}          & \multicolumn{1}{l|}{3}                                                              & 0.75                               \\ \cline{3-9} 
                                                              &                                                      & CWE-416: Use After Free                                                          & \multicolumn{1}{l|}{18}         & \multicolumn{1}{l|}{1}                                                              & 0.06                               & \multicolumn{1}{l|}{4}          & \multicolumn{1}{l|}{0}                                                              & 0.0                                \\ \cline{3-9} 
                                                              &                                                      & CWE-476: NULL Pointer Dereference                                                & \multicolumn{1}{l|}{18}         & \multicolumn{1}{l|}{7}                                                              & 0.39                               & \multicolumn{1}{l|}{4}          & \multicolumn{1}{l|}{2}                                                              & 0.5                                \\ \cline{3-9} 
                                                              &                                                      & CWE-787: Out-of-bounds Write                                                     & \multicolumn{1}{l|}{18}         & \multicolumn{1}{l|}{5}                                                              & 0.28                               & \multicolumn{1}{l|}{4}          & \multicolumn{1}{l|}{2}                                                              & 0.5                                \\ \hline
\multirow{10}{*}{list\_print}                                 & \multirow{3}{*}{Control}                             & CWE-119: Improper Restriction of Operations within the Bounds of a Memory Buffer & \multicolumn{1}{l|}{24}         & \multicolumn{1}{l|}{12}                                                             & 0.5                                & \multicolumn{1}{l|}{9}          & \multicolumn{1}{l|}{2}                                                              & 0.22                               \\ \cline{3-9} 
                                                              &                                                      & CWE-400: Uncontrolled Resource Consumption                                       & \multicolumn{1}{l|}{24}         & \multicolumn{1}{l|}{2}                                                              & 0.08                               & \multicolumn{1}{l|}{9}          & \multicolumn{1}{l|}{1}                                                              & 0.11                               \\ \cline{3-9} 
                                                              &                                                      & CWE-476: NULL Pointer Dereference                                                & \multicolumn{1}{l|}{24}         & \multicolumn{1}{l|}{2}                                                              & 0.08                               & \multicolumn{1}{l|}{9}          & \multicolumn{1}{l|}{1}                                                              & 0.11                               \\ \cline{2-9} 
                                                              & \multirow{5}{*}{Assisted}                            & CWE-119: Improper Restriction of Operations within the Bounds of a Memory Buffer & \multicolumn{1}{l|}{27}         & \multicolumn{1}{l|}{17}                                                             & 0.63                               & \multicolumn{1}{l|}{13}         & \multicolumn{1}{l|}{8}                                                              & 0.62                               \\ \cline{3-9} 
                                                              &                                                      & CWE-400: Uncontrolled Resource Consumption                                       & \multicolumn{1}{l|}{27}         & \multicolumn{1}{l|}{2}                                                              & 0.07                               & \multicolumn{1}{l|}{13}         & \multicolumn{1}{l|}{1}                                                              & 0.08                               \\ \cline{3-9} 
                                                              &                                                      & CWE-416: Use After Free                                                          & \multicolumn{1}{l|}{27}         & \multicolumn{1}{l|}{1}                                                              & 0.04                               & \multicolumn{1}{l|}{13}         & \multicolumn{1}{l|}{1}                                                              & 0.08                               \\ \cline{3-9} 
                                                              &                                                      & CWE-476: NULL Pointer Dereference                                                & \multicolumn{1}{l|}{27}         & \multicolumn{1}{l|}{1}                                                              & 0.04                               & \multicolumn{1}{l|}{13}         & \multicolumn{1}{l|}{0}                                                              & 0.0                                \\ \cline{3-9} 
                                                              &                                                      & CWE-787: Out-of-bounds Write                                                     & \multicolumn{1}{l|}{27}         & \multicolumn{1}{l|}{1}                                                              & 0.04                               & \multicolumn{1}{l|}{13}         & \multicolumn{1}{l|}{0}                                                              & 0.0                                \\ \cline{2-9} 
                                                              & \multirow{2}{*}{Autopilot}                           & CWE-119: Improper Restriction of Operations within the Bounds of a Memory Buffer & \multicolumn{1}{l|}{5}          & \multicolumn{1}{l|}{1}                                                              & 0.2                                & \multicolumn{1}{l|}{1}          & \multicolumn{1}{l|}{1}                                                              & 1.0                                \\ \cline{3-9} 
                                                              &                                                      & CWE-476: NULL Pointer Dereference                                                & \multicolumn{1}{l|}{5}          & \multicolumn{1}{l|}{1}                                                              & 0.2                                & \multicolumn{1}{l|}{1}          & \multicolumn{1}{l|}{0}                                                              & 0.0                                \\ \hline
\multirow{8}{*}{list\_remove\_item\_at\_pos}                  & \multirow{2}{*}{Control}                             & CWE-400: Uncontrolled Resource Consumption                                       & \multicolumn{1}{l|}{13}         & \multicolumn{1}{l|}{28}                                                             & 2.15                               & \multicolumn{1}{l|}{10}         & \multicolumn{1}{l|}{25}                                                             & 2.5                                \\ \cline{3-9} 
                                                              &                                                      & CWE-476: NULL Pointer Dereference                                                & \multicolumn{1}{l|}{13}         & \multicolumn{1}{l|}{21}                                                             & 1.62                               & \multicolumn{1}{l|}{10}         & \multicolumn{1}{l|}{18}                                                             & 1.8                                \\ \cline{2-9} 
                                                              & \multirow{4}{*}{Assisted}                            & CWE-119: Improper Restriction of Operations within the Bounds of a Memory Buffer & \multicolumn{1}{l|}{19}         & \multicolumn{1}{l|}{1}                                                              & 0.05                               & \multicolumn{1}{l|}{13}         & \multicolumn{1}{l|}{0}                                                              & 0.0                                \\ \cline{3-9} 
                                                              &                                                      & CWE-190: Integer Overflow or Wraparound                                          & \multicolumn{1}{l|}{19}         & \multicolumn{1}{l|}{1}                                                              & 0.05                               & \multicolumn{1}{l|}{13}         & \multicolumn{1}{l|}{0}                                                              & 0.0                                \\ \cline{3-9} 
                                                              &                                                      & CWE-400: Uncontrolled Resource Consumption                                       & \multicolumn{1}{l|}{19}         & \multicolumn{1}{l|}{47}                                                             & 2.47                               & \multicolumn{1}{l|}{13}         & \multicolumn{1}{l|}{36}                                                             & 2.77                               \\ \cline{3-9} 
                                                              &                                                      & CWE-476: NULL Pointer Dereference                                                & \multicolumn{1}{l|}{19}         & \multicolumn{1}{l|}{25}                                                             & 1.32                               & \multicolumn{1}{l|}{13}         & \multicolumn{1}{l|}{15}                                                             & 1.15                               \\ \cline{2-9} 
                                                              & \multirow{2}{*}{Autopilot}                           & CWE-400: Uncontrolled Resource Consumption                                       & \multicolumn{1}{l|}{5}          & \multicolumn{1}{l|}{7}                                                              & 1.4                                & \multicolumn{1}{l|}{3}          & \multicolumn{1}{l|}{3}                                                              & 1.0                                \\ \cline{3-9} 
                                                              &                                                      & CWE-476: NULL Pointer Dereference                                                & \multicolumn{1}{l|}{5}          & \multicolumn{1}{l|}{9}                                                              & 1.8                                & \multicolumn{1}{l|}{3}          & \multicolumn{1}{l|}{6}                                                              & 2.0                                \\ \hline
\multirow{5}{*}{list\_save}                                   & \multirow{3}{*}{Control}                             & CWE-119: Improper Restriction of Operations within the Bounds of a Memory Buffer & \multicolumn{1}{l|}{14}         & \multicolumn{1}{l|}{1}                                                              & 0.07                               & \multicolumn{1}{l|}{7}          & \multicolumn{1}{l|}{0}                                                              & 0.0                                \\ \cline{3-9} 
                                                              &                                                      & CWE-400: Uncontrolled Resource Consumption                                       & \multicolumn{1}{l|}{14}         & \multicolumn{1}{l|}{2}                                                              & 0.14                               & \multicolumn{1}{l|}{7}          & \multicolumn{1}{l|}{0}                                                              & 0.0                                \\ \cline{3-9} 
                                                              &                                                      & CWE-787: Out-of-bounds Write                                                     & \multicolumn{1}{l|}{14}         & \multicolumn{1}{l|}{1}                                                              & 0.07                               & \multicolumn{1}{l|}{7}          & \multicolumn{1}{l|}{1}                                                              & 0.14                               \\ \cline{2-9} 
                                                              & \multirow{2}{*}{Assisted}                            & CWE-400: Uncontrolled Resource Consumption                                       & \multicolumn{1}{l|}{17}         & \multicolumn{1}{l|}{3}                                                              & 0.18                               & \multicolumn{1}{l|}{7}          & \multicolumn{1}{l|}{1}                                                              & 0.14                               \\ \cline{3-9} 
                                                              &                                                      & CWE-787: Out-of-bounds Write                                                     & \multicolumn{1}{l|}{17}         & \multicolumn{1}{l|}{2}                                                              & 0.12                               & \multicolumn{1}{l|}{7}          & \multicolumn{1}{l|}{1}                                                              & 0.14                               \\ \hline
\multirow{9}{*}{list\_swap\_item\_positions}                  & \multirow{3}{*}{Control}                             & CWE-119: Improper Restriction of Operations within the Bounds of a Memory Buffer & \multicolumn{1}{l|}{12}         & \multicolumn{1}{l|}{1}                                                              & 0.08                               & \multicolumn{1}{l|}{5}          & \multicolumn{1}{l|}{0}                                                              & 0.0                                \\ \cline{3-9} 
                                                              &                                                      & CWE-400: Uncontrolled Resource Consumption                                       & \multicolumn{1}{l|}{12}         & \multicolumn{1}{l|}{2}                                                              & 0.17                               & \multicolumn{1}{l|}{5}          & \multicolumn{1}{l|}{0}                                                              & 0.0                                \\ \cline{3-9} 
                                                              &                                                      & CWE-476: NULL Pointer Dereference                                                & \multicolumn{1}{l|}{12}         & \multicolumn{1}{l|}{23}                                                             & 1.92                               & \multicolumn{1}{l|}{5}          & \multicolumn{1}{l|}{11}                                                             & 2.2                                \\ \cline{2-9} 
                                                              & \multirow{4}{*}{Assisted}                            & CWE-119: Improper Restriction of Operations within the Bounds of a Memory Buffer & \multicolumn{1}{l|}{20}         & \multicolumn{1}{l|}{3}                                                              & 0.15                               & \multicolumn{1}{l|}{6}          & \multicolumn{1}{l|}{0}                                                              & 0.0                                \\ \cline{3-9} 
                                                              &                                                      & CWE-190: Integer Overflow or Wraparound                                          & \multicolumn{1}{l|}{20}         & \multicolumn{1}{l|}{2}                                                              & 0.1                                & \multicolumn{1}{l|}{6}          & \multicolumn{1}{l|}{0}                                                              & 0.0                                \\ \cline{3-9} 
                                                              &                                                      & CWE-400: Uncontrolled Resource Consumption                                       & \multicolumn{1}{l|}{20}         & \multicolumn{1}{l|}{9}                                                              & 0.45                               & \multicolumn{1}{l|}{6}          & \multicolumn{1}{l|}{0}                                                              & 0.0                                \\ \cline{3-9} 
                                                              &                                                      & CWE-476: NULL Pointer Dereference                                                & \multicolumn{1}{l|}{20}         & \multicolumn{1}{l|}{29}                                                             & 1.45                               & \multicolumn{1}{l|}{6}          & \multicolumn{1}{l|}{7}                                                              & 1.17                               \\ \cline{2-9} 
                                                              & \multirow{2}{*}{Autopilot}                           & CWE-400: Uncontrolled Resource Consumption                                       & \multicolumn{1}{l|}{5}          & \multicolumn{1}{l|}{4}                                                              & 0.8                                & \multicolumn{1}{l|}{0}          & \multicolumn{1}{l|}{0}                                                              & \textit{}                          \\ \cline{3-9} 
                                                              &                                                      & CWE-476: NULL Pointer Dereference                                                & \multicolumn{1}{l|}{5}          & \multicolumn{1}{l|}{11}                                                             & 2.2                                & \multicolumn{1}{l|}{0}          & \multicolumn{1}{l|}{0}                                                              & \textit{}                          \\ \hline
\multirow{12}{*}{list\_update\_item\_at\_pos}                 & \multirow{4}{*}{Control}                             & CWE-119: Improper Restriction of Operations within the Bounds of a Memory Buffer & \multicolumn{1}{l|}{14}         & \multicolumn{1}{l|}{1}                                                              & 0.07                               & \multicolumn{1}{l|}{10}         & \multicolumn{1}{l|}{0}                                                              & 0.0                                \\ \cline{3-9} 
                                                              &                                                      & CWE-400: Uncontrolled Resource Consumption                                       & \multicolumn{1}{l|}{14}         & \multicolumn{1}{l|}{10}                                                             & 0.71                               & \multicolumn{1}{l|}{10}         & \multicolumn{1}{l|}{10}                                                             & 1.0                                \\ \cline{3-9} 
                                                              &                                                      & CWE-416: Use After Free                                                          & \multicolumn{1}{l|}{14}         & \multicolumn{1}{l|}{10}                                                             & 0.71                               & \multicolumn{1}{l|}{10}         & \multicolumn{1}{l|}{10}                                                             & 1.0                                \\ \cline{3-9} 
                                                              &                                                      & CWE-476: NULL Pointer Dereference                                                & \multicolumn{1}{l|}{14}         & \multicolumn{1}{l|}{17}                                                             & 1.21                               & \multicolumn{1}{l|}{10}         & \multicolumn{1}{l|}{16}                                                             & 1.6                                \\ \cline{2-9} 
                                                              & \multirow{5}{*}{Assisted}                            & CWE-119: Improper Restriction of Operations within the Bounds of a Memory Buffer & \multicolumn{1}{l|}{24}         & \multicolumn{1}{l|}{5}                                                              & 0.21                               & \multicolumn{1}{l|}{16}         & \multicolumn{1}{l|}{1}                                                              & 0.06                               \\ \cline{3-9} 
                                                              &                                                      & CWE-400: Uncontrolled Resource Consumption                                       & \multicolumn{1}{l|}{24}         & \multicolumn{1}{l|}{26}                                                             & 1.08                               & \multicolumn{1}{l|}{16}         & \multicolumn{1}{l|}{22}                                                             & 1.38                               \\ \cline{3-9} 
                                                              &                                                      & CWE-416: Use After Free                                                          & \multicolumn{1}{l|}{24}         & \multicolumn{1}{l|}{22}                                                             & 0.92                               & \multicolumn{1}{l|}{16}         & \multicolumn{1}{l|}{19}                                                             & 1.19                               \\ \cline{3-9} 
                                                              &                                                      & CWE-476: NULL Pointer Dereference                                                & \multicolumn{1}{l|}{24}         & \multicolumn{1}{l|}{30}                                                             & 1.25                               & \multicolumn{1}{l|}{16}         & \multicolumn{1}{l|}{18}                                                             & 1.13                               \\ \cline{3-9} 
                                                              &                                                      & CWE-787: Out-of-bounds Write                                                     & \multicolumn{1}{l|}{24}         & \multicolumn{1}{l|}{5}                                                              & 0.21                               & \multicolumn{1}{l|}{16}         & \multicolumn{1}{l|}{0}                                                              & 0.0                                \\ \cline{2-9} 
                                                              & \multirow{3}{*}{Autopilot}                           & CWE-400: Uncontrolled Resource Consumption                                       & \multicolumn{1}{l|}{5}          & \multicolumn{1}{l|}{6}                                                              & 1.2                                & \multicolumn{1}{l|}{3}          & \multicolumn{1}{l|}{5}                                                              & 1.67                               \\ \cline{3-9} 
                                                              &                                                      & CWE-416: Use After Free                                                          & \multicolumn{1}{l|}{5}          & \multicolumn{1}{l|}{3}                                                              & 0.6                                & \multicolumn{1}{l|}{3}          & \multicolumn{1}{l|}{2}                                                              & 0.67                               \\ \cline{3-9} 
                                                              &                                                      & CWE-476: NULL Pointer Dereference                                                & \multicolumn{1}{l|}{5}          & \multicolumn{1}{l|}{7}                                                              & 1.4                                & \multicolumn{1}{l|}{3}          & \multicolumn{1}{l|}{6}                                                              & 2.0                                \\ \hline
\end{tabular}